\documentclass[twocolumn,showpacs,preprintnumbers,amsmath,amssymb,
 aps,
 prb,
 lengthcheck,%
]{revtex4-1}
\usepackage{amssymb}
\usepackage{graphicx}
\usepackage{dcolumn}
\usepackage{bm}
\usepackage{hyperref}
\usepackage{makeidx}
\usepackage[english]{babel} 
\usepackage[latin1]{inputenc}
\usepackage{color}

\usepackage{latexsym}
\usepackage{amsthm}
\usepackage{epstopdf}
\usepackage{dsfont}

\begin{document}

\title{A proposal for the observation of nonlocal multipair production: the biSQUID}

\author{J. Rech$^1$, T. Jonckheere$^1$, T. Martin$^1$,
  B. Dou\c{c}ot$^2$, D. Feinberg$^{3,4}$, R. M\'elin$^{3,4}$} 

\affiliation{$^1$ Aix Marseille Universit\'e, Universit\'e de Toulon, CNRS, CPT, UMR 7332, 13288 Marseille, France}
\affiliation{$^2$ Laboratoire de Physique Th\'eorique et des Hautes Energies,
  CNRS UMR 7589, Universit\'es Paris 6 et 7, 4 Place Jussieu, 75252 Paris
  Cedex 05}
\affiliation{$^3$ Centre National de la Recherche Scientifique, Institut NEEL, F-38042 Grenoble Cedex 9, France}
\affiliation{$^4$ Universit\'e Grenoble-Alpes, Institut NEEL, F-38042 Grenoble Cedex 9, France}

\date{\today}

\begin{abstract}
We propose an all-superconducting three-terminal setup consisting in a carbon nanotube (or semiconducting nanowire) contacted to three superconducting leads. The resulting device, referred to as a "biSQUID", is made of four quantum dots arranged in two loops of different surface area. We show how this biSQUID can prove a useful tool to probe nonlocal quantum phenomena in an interferometry setup.
We study the measured critical current as a function of the applied magnetic field, which shows peaks in its Fourier spectrum, providing clear signatures of multipair Josephson processes. The device does not require any specific fine-tuning as these features are observed for a wide range of microscopic parameters -- albeit with a non-trivial dependence. Competing effects which may play a significant role in actual experimental realizations are also explored.
\end{abstract}


\maketitle


\section{Introduction} \label{sec:introduction}

Multiterminal setups offer a great way to explore nonlocal quantum effects as well as entanglement in condensed matter devices, attracting both experimental and theoretical attention. Recently, these efforts have focused on all-superconducting hybrid structures, involving quantum dots or metallic islands connected to multiple superconducting leads.\cite{cuevas2007,duhot2009,houzet2010,chtchelkatchev2010,kaviraj2011} In particular, the prediction of nonlocal quartet production,\cite{freyn2011} that is the emission of spatially correlated pairs of Cooper pairs, opens new perspectives in the realization of electronic entanglers. 

Motivated by this, a recent work \cite{jonckheere2013} considered an all-superconducting bijunction consisting of a central superconductor coupled via gate-controllable quantum dots to two lateral voltage-biased superconductors. It uncovered the presence of a coherent transport mechanism away from equilibrium, manifesting as multipair phase-coherent Josephson resonances in the current, appearing on top of the usual local dissipative transport of quasiparticles.
Similarly, experimental work performed on a three-terminal voltage-biased Josephson junction involving a central T-shaped metallic region revealed features in the electronic subgap transport consistent with the production of nonlocal quartets \cite{pfeffer2013} though an unequivocal signature is still lacking at the moment.
In this context, alternative ways of detecting such nonlocal multipair processes are highly desirable, in particular in the coherent dissipationless regime, a route that could be provided by an interferometric setup, such as a SQUID.


\begin{figure}[tb]
\centering
\includegraphics[scale=0.19]{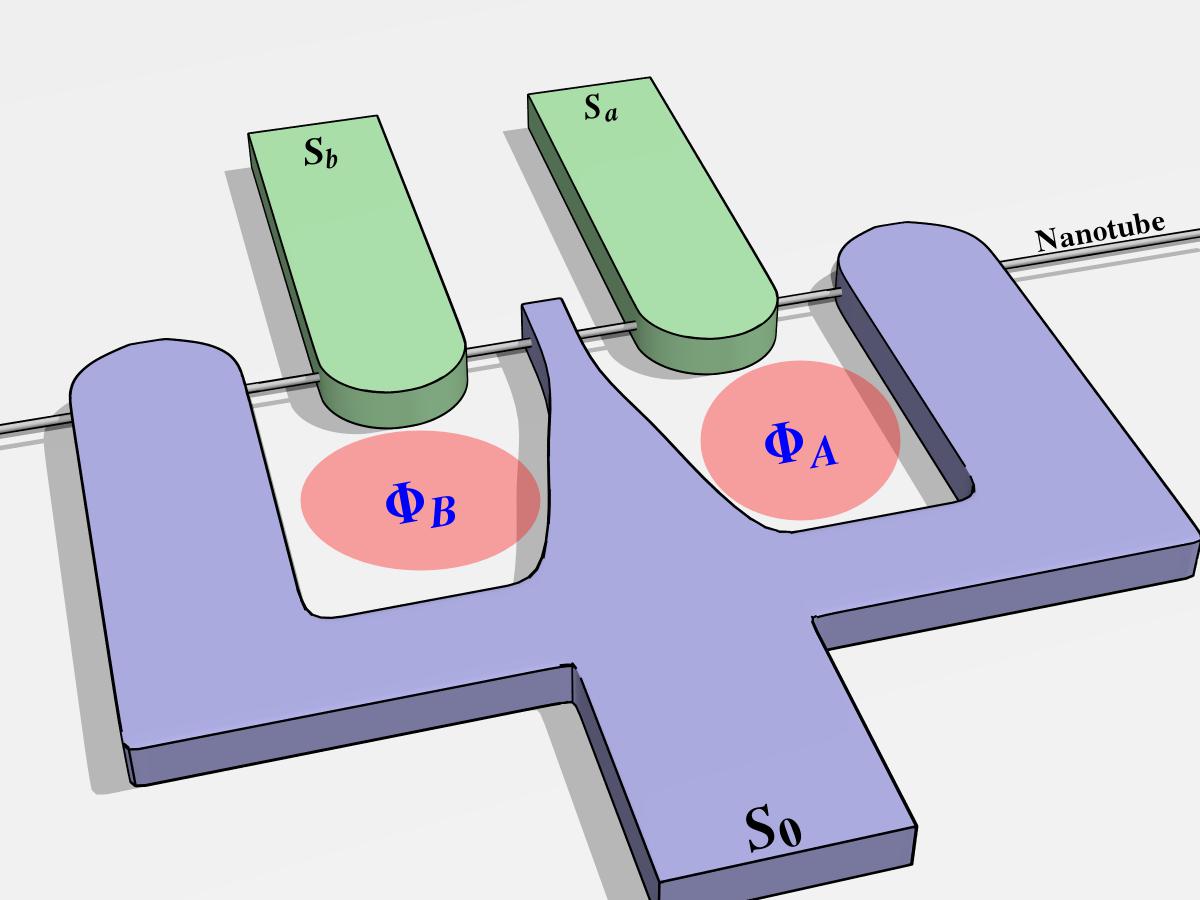}
\caption{Artistic view of the biSQUID setup where a single nanotube or nanowire is contacted to three superconducting electrodes (all at the same potential), with magnetic field biasing. 
}
\label{fig:reallife}
\end{figure}


A superconductor quantum interference device, or SQUID, consists in a superconducting loop, defining two paths each interrupted by  a Josephson junction. As a direct consequence of phase coherence, the Cooper pairs flowing along these two paths interfere, in a way that is controlled by the magnetic flux through the loop. 
In a recent achievement of molecular electronics, a carbon nanotube SQUID has been realized experimentally. \cite{cleuziou2006} There, the two constitutive Josephson junctions are both made of a nanotube quantum dot, allowing to tune their transparency with the help of an external gate voltage, therefore providing a new generation of versatile sensors.  
Interestingly, similar carbon nanotube (or nanowire) devices have been used as Cooper pair beam-splitters, \cite{hofstetter2009,herrmann2010,burset2011,chevallier2011,das2012} a source of entangled electron pairs\cite{lesovik2001,recher2001,samuelsson2003} where the quasi one-dimensional nanostructure is contacted to a central superconducting source and two metallic drains. 
These rely on a specific nonlocal process, referred to as CAR for crossed Andreev reflection,\cite{byers1995,martin1996,anantram1996,deutscher2000,melin2001} which amounts to separating the constituents of a Cooper pair into spatially distinct contacts (provided they are within a distance set by the coherence length).

It is therefore only natural to devise a setup bringing together the interferometric properties of the SQUID with the nonlocal aspects associated with CAR processes. Such a setup could be constructed from a carbon nanotube with readily accessible technology, as illustrated by the many examples of multiterminal nanotube-based devices now available.\cite{biercuk2005, gao2005, gunnarsson2008, feuilletpalma2010}

In this paper, we propose such a device, dubbed a biSQUID, consisting of two twinned SQUIDs, where the superconducting contact which emits the multiple pairs is common to both loops in order to reveal multipair processes through interferometry. The two loops realize a three-terminal structure made of a single carbon nanotube (or nanowire) contacted to three superconducting leads (see Fig.~\ref{fig:reallife}), delimiting four quantum dots which are controlled by external gate voltages. The device is ideally operated by fixing the total current flowing through, which we express in terms of the magnetic fluxes piercing the two loops.

Recently, a two-terminal SQUID geometry involving two loops was realized experimentally using superconductor - normal metal - superconductor junctions.\cite{ronzani2013}
While somewhat similar in spirit to the present proposal, this setup was specifically designed for sensing applications rather than to explore nonlocal effects.
It shows, however, that not only the actual realization of these systems is within our grasp
but also that such superconducting nanodevices constitute an active field of research.

The outline of the paper is as follows. In section \ref{sec:setup}, we introduce the setup and present a first simple phenomenological approach. We then derive the expression for the critical current in section \ref{sec:micro}, starting from a microscopic description of the setup. Section \ref{sec:critical} is devoted to our results, where we show and analyze the dependence of the critical current on the external magnetic field, and comment on the robustness of the observed features. Finally, in section \ref{sec:competing}, we explore the competing effects which might spoil the expected signatures of pure CAR processes, before concluding in section \ref{sec:conclusion}.


\section{Setup and phenomenology} \label{sec:setup}

\begin{figure}[tb]
\centering
\includegraphics[scale=1]{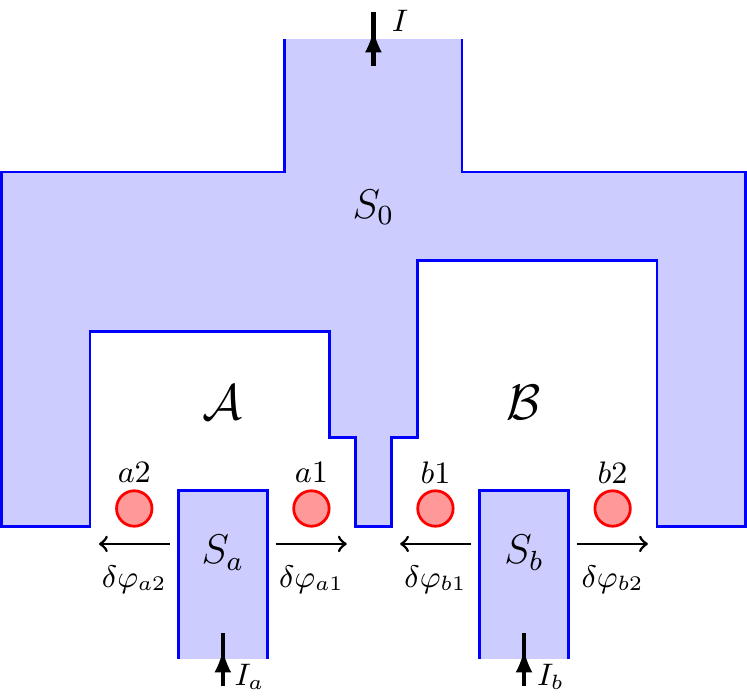}
\caption{The biSQUID setup as modeled here. Three superconducting terminals and four quantum dots (originating from a single nanotube or nanowire) define a two-loop system with enclosed areas of different sizes. 
Each dot Josephson junction sees a phase difference labeled $\delta \varphi_{\alpha}$ combining the phases of each superconducting electrodes as well as the enclosed magnetic fluxes. The arrows near each junction set the conventions for the flowing currents and phase differences.
}
\label{fig:setup}
\end{figure}

The biSQUID setup is presented in Fig.~\ref{fig:setup}. 
It basically amounts to twinning two nanotube-based SQUIDs by a common central electrode whose width is smaller than the coherence length, in order to support nonlocal Andreev scattering processes.\footnote{For simplicity, the other two electrodes $S_a$ and $S_b$ are chosen wide enough so as to avoid nonlocal effects.}
A carbon nanotube (or nanowire) is contacted with three superconducting electrodes, referred to as $S_a$, $S_0$ and $S_b$. This defines four quantum dots, labeled $a1$, $a2$, $b1$ and $b2$, controllable via nearby electrostatic gates. The area of the two resulting SQUID loops (denoted $\mathcal{A}$ and $\mathcal{B}$ respectively) are chosen different, so as to ensure that electrons feel different magnetic fluxes depending on the loop they flow through.  

In order to make the upcoming discussion as clear as possible, it is important to properly define the various superconducting phases involved in this setup. A phase $\varphi_s$ is attributed to the central superconducting finger, while the other two superconducting electrodes are characterized respectively by a phase $\varphi_a$ and $\varphi_b$. Furthermore, the magnetic fluxes through the two loops $\mathcal{A}$ and $\mathcal{B}$ are defined as $\Phi_\mathcal{A} = B \mathcal{S}_\mathcal{A}$ and $\Phi_\mathcal{B} = B \mathcal{S}_\mathcal{B}$, where $\mathcal{S}_j$ is the surface area enclosed by loop $j$. 

It follows from flux quantization that the phase differences seen by each quantum dot Josephson junction, are given by
\begin{align}
\label{eq:deltaphia1}
\delta \varphi_{a1} &= \varphi_s - \varphi_a \\ 
\delta \varphi_{b1} &= \varphi_s - \varphi_b \\ 
\delta \varphi_{a2} &= \varphi_s - \varphi_a - 2 \pi \frac{\Phi_{\mathcal{A}}}{\Phi_0} \\ 
\delta \varphi_{b2} &= \varphi_s - \varphi_b + 2 \pi \frac{\Phi_{\mathcal{B}}}{\Phi_0} ,
\label{eq:deltaphib2}
\end{align}
where $\Phi_0 = h/ (2 e)$ is the flux quantum. The biSQUID is controlled by fixing the total current $I = I_a + I_b$ measured from the common output, which depends on all of these four phase differences, or alternatively on the superconducting phases $\varphi_a$ and $\varphi_b$ as well as the magnetic fluxes through the loops $\Phi_{\mathcal{A}}$ and $\Phi_{\mathcal{B}}$ (from this point on, we set $\varphi_s =0$ as the reference for the superconducting phases). 

A key quantity of interest for such a superconducting interferometer is the flux-dependent critical current, i.e. the maximum dissipationless current which can flow through the device, defined here as
\begin{equation}
I_c \left( \Phi_{\mathcal{A}} , \Phi_{\mathcal{B}} \right) = \underset{\varphi_a, \varphi_b}{\text{Max}} \left| I \left( \delta \varphi_{a1}, \delta \varphi_{b1}, \delta \varphi_{a2}, \delta \varphi_{b2}\right) \right| .
\end{equation}
Out of the four quantum-dot based junctions, only $a1$ and $b1$ are coupled by nonlocal effects, leading to signatures in the critical current involving specific combinations of the external fluxes, which can be revealed through a simple phenomenological approach.

In the limit of low transparency, we can perform a simple perturbative treatment in the Cooper pair tunneling. 
Processes involving a single pair contribute to the total current $I$ measured from $S_0$ in the standard Josephson form of a sinusoidal current-phase relationship for each of the four junctions, leading to
\begin{equation}
I_{1\text{P}} = I_{\text{J}} \left[ \sin \delta\varphi_{a1} + \sin \delta\varphi_{b1} + \sin \delta\varphi_{a2} + \sin \delta\varphi_{b2} \right].
\end{equation}
Processes involving two pairs all lead to second-order harmonics of the Josephson current. However, one must distinguish between two different contributions. 
Local processes, amounting to two pairs crossing any given junction, involve twice the phase difference seen by each quantum dot:
\begin{align}
I_{2\text{P,local}} = I_{\text{JJ}} & \left[ \sin \left(2 \delta\varphi_{a1}\right) + \sin \left( 2 \delta\varphi_{b1} \right) \right. \nonumber \\
 & \left. + \sin \left( 2 \delta\varphi_{a2}\right) + \sin \left( 2 \delta\varphi_{b2} \right) \right].
\end{align}
In addition to these, the present device allows the possibility for nonlocal processes, and one thus needs to consider the so-called pair cotunneling and quartet supercurrents.\cite{freyn2011} The latter corresponds to the splitting of two correlated pairs from $S_0$ into $S_a$ and $S_b$, which leads to a Josephson-like current-phase relationship involving both the phase difference between $S_0$ and $S_a$, and the one between $S_0$ and $S_b$, namely
\begin{align}
I_{2\text{P,quartet}} = I_{\text{Q}}  \sin \left( \delta\varphi_{a1} + \delta\varphi_{b1} \right). 
\label{eq:quartetcurrent}
\end{align}
Due to the exchange process intrinsic to quartet emission, we expect $I_{\text{Q}}$ to be negative.\cite{jonckheere2013} This nonlocal process relies on crossed Andreev reflection through $S_0$, which naturally coexists with normal transmission through the central superconducting electrode without electron-hole conversion, the so-called elastic cotunneling. It follows that the quartet process comes with a similar partner corresponding to the exchange of a pair from $S_a$ to $S_b$ via double elastic cotunneling through $S_0$, therefore contributing to the supercurrent as
\begin{align}
I_{2\text{P,pair cotunneling}} = I_{\text{PC}}  \sin \left( \delta\varphi_{a1} - \delta\varphi_{b1} \right),
\label{eq:paircotunnelingcurrent}
\end{align}
where the various current scales introduced above satisfy $|I_\text{PC}|, |I_{\text{Q}}|, |I_{\text{JJ}}| 	\ll I_{\text{J}}$, as a consequence of the low transparency of the junctions.

Combining Eqs.~\eqref{eq:deltaphia1} though \eqref{eq:paircotunnelingcurrent}, one obtains the following expression for the total critical current, up to second order in the pair tunneling
\begin{align}
I_c \left( \Phi_{\mathcal{A}} , \Phi_{\mathcal{B}} \right) = \underset{\varphi_a, \varphi_b}{\text{Max}}  
&\left|
2 I_{\text{J}}  \sin \left( \varphi_a + \pi \tilde{\Phi}_\mathcal{A} \right)  \cos \left( \pi \tilde{\Phi}_\mathcal{A} \right) \right. \nonumber \\
& + 2 I_{\text{J}} \sin \left( \varphi_b - \pi \tilde{\Phi}_\mathcal{B} \right)  \cos \left( \pi \tilde{\Phi}_\mathcal{B} \right) 
\nonumber \\
& + 2 I_{\text{JJ}} \sin \left( 2 \varphi_a + 2 \pi \tilde{\Phi}_\mathcal{A} \right)  \cos \left( 2 \pi \tilde{\Phi}_\mathcal{A} \right) 
\nonumber \\
& + 2 I_{\text{JJ}} \sin \left( 2 \varphi_b - 2 \pi \tilde{\Phi}_\mathcal{B} \right)  \cos \left( 2 \pi \tilde{\Phi}_\mathcal{B} \right)
\nonumber \\
& + I_{\text{Q}}  \sin \left( \varphi_{a} + \varphi_{b} \right) 
\nonumber \\
& + I_{\text{PC}}  \sin \left( \varphi_{a} - \varphi_{b} \right)
\Big|
\label{eq:maxphenom}
\end{align}
where $\tilde{\Phi}_j = \Phi_j / \Phi_0$ is the magnetic flux through loop $j$, in units of the flux quantum.
Apart from small regions near integer values of $\tilde{\Phi}_{\mathcal{A},\mathcal{B}}$ (where the single pair Josephson current is near suppression), the critical current can readily be obtained without any further calculation in most of the $(\Phi_\mathcal{A},\Phi_\mathcal{B})$-plane, and takes the form
\begin{align}
I_c \left( \Phi_{\mathcal{A}} , \Phi_{\mathcal{B}} \right) =& 2  I_{\text{J}} \left[ \left| \cos \left( \pi \tilde{\Phi}_\mathcal{A} \right) \right| + \left| \cos \left( \pi \tilde{\Phi}_\mathcal{B} \right) \right| \right] 
\nonumber \\
& + |I_{\text{Q}}| \left| \sin \left( \pi \tilde{\Phi}_\mathcal{A} - \pi \tilde{\Phi}_\mathcal{B} \right) \right| 
\nonumber \\
& + |I_{\text{PC}}| \left| \sin \left( \pi \tilde{\Phi}_\mathcal{A} + \pi \tilde{\Phi}_\mathcal{B} \right) \right| .
\label{eq:current-phenom}
\end{align}
There, the first term corresponds to the critical current in the absence of a quartet supercurrent. The two loops are decoupled in this case and $I_c \left( \Phi_{\mathcal{A}} , \Phi_{\mathcal{B}} \right)$ splits into two independent contributions corresponding to the simultaneous maximization of $I_a$ and $I_b$. 
Interestingly, quartet and pair cotunneling processes are responsible for an extra contribution to the critical current with a very specific flux dependence, leading to a macroscopic manifestation of nonlocal effects.
In particular, this simple calculation points out that a measurement of the quartet current is possible by detecting the flux periodicity of the biSQUID critical current.



\section{Microscopic theory} \label{sec:micro}

This section is devoted to a microscopic calculation of the current through the biSQUID. Our goal is to justify the expression \eqref{eq:current-phenom} obtained from our simple phenomenological approach, to generalize it to more transparent junctions, and to show that the energy levels in the dots can be chosen such as to promote multi-pair transport, and in particular reveal the presence of a quartet resonance.

\subsection{Hamiltonian} \label{sec:hamilt}

The model Hamiltonian for the biSQUID setup is expressed as the sum of three contributions
\begin{equation}
\mathcal{H} = \sum_{j=a,0,b} \mathcal{H}_{S,j} + \sum_{\substack{\alpha=a1,a2,\\ b1,b2}}  \mathcal{H}_{D,\alpha} + \mathcal{H}_T .
\end{equation}
Here $\mathcal{H}_{S,j}$ is the Hamiltonian associated with the superconducting lead $S_j$ ($j=a,0,b$), which is given by the following compact form
\begin{equation}
\mathcal{H}_{S,j} = \sum_k \Psi^\dagger_{jk} \left(
\xi_k \sigma_z + \Delta_j  \sigma_x \right) \Psi_{jk} ,
\end{equation}
where $\xi_k = \frac{k^2}{2 m} - \mu$ and $\Delta_j$ is the superconducting gap of lead $j$. 
Each quantum dot $\alpha$ ($\alpha=a1,a2,b1,b2$) is modeled by a single non-interacting level, with energy $\epsilon_\alpha$, described by the Hamiltonian
\begin{equation}
\mathcal{H}_{D,\alpha} = \epsilon_\alpha {\bf d}^\dagger_{\alpha}  \sigma_z  {\bf d}_{\alpha} .
\end{equation}
In both these expressions, we used Pauli matrices acting in Nambu space, and introduced Nambu spinors for the lead and dot electrons, defined respectively as
\begin{equation}
\Psi_{jk} = 
\begin{pmatrix}
\psi_{j, k, \uparrow} \\
\psi^\dagger_{j, -k, \downarrow}
\end{pmatrix}
\qquad 
{\bf d}_{\alpha} =
\begin{pmatrix}
d_{\alpha \uparrow} \\
d^\dagger_{\alpha \downarrow}
\end{pmatrix} .
\end{equation}
Introducing a tunneling amplitude $t_{j\alpha}$ between lead $j$ and dot $\alpha$, and performing a gauge transformation to incorporate the superconducting phases $\varphi_j$ in the tunneling term, one has for the tunneling part of the Hamiltonian
\begin{equation}
\mathcal{H}_T = \sum_k \left( \Psi^\dagger_{jk} \mathcal{T}_{jk \alpha} {\bf d}_\alpha + \text{H.c.} \right)
\end{equation}
where, in absence of external magnetic field, we have $\mathcal{T}_{jk\alpha} = t_{j \alpha} e^{-i k r_{j\alpha}} \sigma_z e^{i \sigma_z \varphi_j /2}$, $r_{j\alpha}$ being the position of lead $j$ where tunneling to/from dot $\alpha$ occurs.

The resulting Hamiltonian is fully quadratic in terms of both the leads and the dot electrons. It is therefore convenient to integrate out the leads degrees of freedom and derive an effective theory involving only the dot electrons.

This is achieved through standard techniques, starting from the partition function
\begin{align}
Z = \int \mathcal{D} \left[ \bar{\psi},\psi , \bar{d}, d\right] e^{- S \left[ \bar{\psi},\psi , \bar{d}, d \right]} ,
\end{align}
with action
\begin{align}
S &= \int_0^\beta d\tau  \sum_{j,k} \bar{\Psi}_{jk}(\tau) \left( D_\tau + 
\xi_k \sigma_z + \Delta_j  \sigma_x \right) \Psi_{jk}(\tau) \nonumber \\
&+\int_0^\beta d\tau \sum_\alpha \bar{{\bf d}}_{\alpha}(\tau) \left( D_\tau + \epsilon_\alpha  \sigma_z  \right) {\bf d}_{\alpha}(\tau) \nonumber \\
&+\int_0^\beta d\tau \sum_{j,\alpha,k}  \left( \bar{\Psi}_{jk}(\tau) \mathcal{T}_{jk \alpha} {\bf d}_\alpha (\tau) +  \bar{{\bf d}}_\alpha (\tau) \mathcal{T}^*_{jk \alpha} \Psi_{jk}(\tau) \right),
\end{align}
where $D_\tau$ is defined in Nambu space as 
$D_\tau = \left( \begin{smallmatrix}
\overset{\rightarrow}{\partial}_\tau & 0 \\
0 & - \overset{\leftarrow}{\partial}_\tau \\
\end{smallmatrix} \right)$.

Carrying out the integration over the $(\bar{\psi},\psi)$ Grassmann fields, one is left with an effective action of the form
\begin{align}
S_\text{eff} =& \int_0^\beta d\tau \sum_\alpha \bar{{\bf d}}_{\alpha}(\tau) \left( D_\tau + \epsilon_\alpha  \sigma_z  \right) {\bf d}_{\alpha}(\tau) \nonumber \\
&+\int_0^\beta d\tau d\tau' \sum_{\alpha,\gamma}  \bar{{\bf d}}_{\alpha}(\tau) \Sigma_{\alpha \gamma} (\tau - \tau') {\bf d}_{\gamma}(\tau')
\end{align}
where we introduced the tunneling self-energy
\begin{align}
\Sigma_{\alpha \gamma} (\tau) = \sum_{j,k} \mathcal{T}^*_{jk \alpha} \mathcal{G}_{j,k} (\tau) \mathcal{T}_{jk \gamma}
\label{eq:tunnelsigma}
\end{align}
which depends on the leads electrons Green's function, defined in Matsubara frequency space as
\begin{align}
\mathcal{G}_{j,k} (i \omega_n) = \left( i\omega_n \mathds{1} - \xi_k \sigma_z - \Delta_j \sigma_x  \right)^{-1} .
\end{align}

It follows that the effective field theory, quadratic in $(\bar{d},d)$, can be described uniquely in terms of the Matsubara Green's function $\hat{G} (i \omega_n)$ for the dot electrons, which takes the form of a $8 \times 8$ matrix in Nambu-dot space given by 
\begin{widetext}
\begin{align}
\hat{G}^{-1} (i \omega_n) = 
\begin{pmatrix}
g_{a2, a} (i\omega_n, \Phi_\mathcal{A}) & 0 & 0 & 0 \\
0 & g_{a1, a} (i\omega_n, 0) & f (i\omega_n) & 0 \\
0 & f (i\omega_n) & g_{b1, b} (i\omega_n, 0) & 0 \\
0 & 0 & 0 & g_{b2, b} (i\omega_n, -\Phi_\mathcal{B}) 
\end{pmatrix} 
\label{eq:inverseG}
\end{align}
where
\begin{align}
g_{\alpha, j} (i\omega_n, \Phi) &= 
\begin{pmatrix}
i\omega_n \left[ 1+\frac{\pi \nu (0)}{\sqrt{\omega_n^2+\Delta^2}} ( t_{0,\alpha}^2 +t_{j,\alpha}^2 )\right] - \epsilon_{\alpha} & \qquad - \Delta e^{i \Phi} (t_{0,\alpha}^2 + t_{j,\alpha}^2 e^{i \delta\varphi_{\alpha}}) \frac{\pi \nu (0)}{\sqrt{\omega_n^2+\Delta^2}}  \\
- \Delta e^{-i \Phi} (t_{0,\alpha}^2 + t_{j,\alpha}^2 e^{-i \delta\varphi_{\alpha}}) \frac{\pi \nu (0)}{\sqrt{\omega_n^2+\Delta^2}}  & \qquad i\omega_n \left[ 1+ \frac{\pi \nu (0)}{\sqrt{\omega_n^2+\Delta^2}} (t_{0,\alpha}^2 + t_{j,\alpha}^2) \right] + \epsilon_{\alpha} \\
\end{pmatrix} \\
f (i\omega_n) &=  \pi \nu(0) t_{0,a1} t_{0,b1} e^{- R / \xi(\omega_n)}
\begin{pmatrix}
 \frac{i \omega_n}{\sqrt{\omega_n^2 + \Delta^2}} \cos (k_F R) - \sin (k_F R) & \quad - \frac{ \Delta }{\sqrt{\omega_n^2 + \Delta^2}} \cos (k_F R)  \\
 - \frac{ \Delta }{\sqrt{\omega_n^2 + \Delta^2}} \cos (k_F R) & \quad \frac{i \omega_n}{\sqrt{\omega_n^2 + \Delta^2}} \cos (k_F R) + \sin (k_F R)  \\
\end{pmatrix}
\label{eq:anomalousGreen}
\end{align}
\end{widetext}
where we introduced the width $R$ of the central superconducting lead, and assumed all superconducting electrodes to have the same gap energy $\Delta_j = \Delta$, and density of states at the Fermi level $\nu(0)$. The energy-dependent coherence length is defined as $\xi (\omega_n) =  \xi_0 \Delta / \sqrt{\omega_n^2 + \Delta^2}$.

\subsection{Current}

The current through a given quantum dot Josephson junction $\alpha$, can be readily expressed as the time derivative of the number of electrons out of the superconducting reservoir, and as such can be related to the phase derivative of the tunneling Hamiltonian. In particular, the average current through dot $\alpha$ reads
\begin{align}
I_\alpha &= \frac{2e}{\hbar} \left\langle \frac{\partial \mathcal{H}_T}{\partial \delta \varphi_\alpha}  \right\rangle = \frac{2e}{\hbar}  \frac{\partial F}{\partial \delta \varphi_\alpha} .
\end{align}
where $F = - k_B T \log Z$ is the free energy.

Keeping in mind that the effective theory contains all the relevant phase-dependent degrees of freedom, this can be further simplified as
\begin{align}
I_\alpha &=  - \frac{2e}{\beta \hbar} \frac{\partial \log Z_\text{eff}}{\partial \delta \varphi_\alpha} \nonumber \\
&=  - \frac{2e}{\beta \hbar} \frac{\partial}{\partial \delta \varphi_\alpha}
 \sum_n \log \left[ \text{det}~ \hat{G}^{-1} (i \omega_n) \right]
\label{eq:currentvsG}
\end{align}
where we performed explicitly the integration over the Grassmann fields $(\bar{d},d)$. The current through any given junction can thus be obtained from Eq.~\eqref{eq:inverseG}, after performing a Matsubara frequency summation. The total current through the device is readily obtained by summing up the contribution coming from each junction.

The width $R$ of the central superconducting lead, which corresponds to the separation between tunneling points from dot $a1$ to $S_0$ and dot $b1$ to $S_0$, enters the above expression for the current in two important ways. First, the total current contains terms which decrease exponentially with $R$. These  correspond to nonlocal contributions which expectedly vanish if the separation between tunneling points exceeds the superconducting coherence length, therefore setting a typical order of magnitude for $R$. Second, some of these terms are also oscillating on a scale set by the Fermi wavelength $\lambda_F$.  Since $\lambda_F$ is typically several orders of magnitude smaller than the superconducting coherence length, these contributions are rapidly oscillating as a function of $R$. It is thus natural to average over these rapid oscillations, assuming that the separation between tunneling points is susceptible to fluctuate slightly on a scale given by a few Fermi wavelengths. 
Defining $R = R_0 + r$, where $R_0$ is a fraction of the coherence length $\xi_0$ and $r$ is of the order of the Fermi wavelength, we introduce the $r$-averaged total current as
\begin{align}
\bar{I}_\text{Tot} = \frac{1}{2 \lambda_F} \int_{-\lambda_F}^{\lambda_F} dr \frac{-2e}{\beta \hbar} \sum_n \sum_\alpha  \frac{\partial  \log \left[ \text{det}~ \hat{G}^{-1} (i \omega_n) \right] }{\partial \delta \varphi_{\alpha}}  .
\label{eq:totalI}
\end{align}
Considering the maximization with respect to the superconducting phases, the critical current through the device is now expressed in terms of the microscopic parameters as
\begin{align}
I_c \left( \Phi_{\mathcal{A}} , \Phi_{\mathcal{B}} \right) = \underset{\varphi_a, \varphi_b}{\text{Max}} &
\left| \bar{I}_\text{Tot} \left( \delta \varphi_{a1}, \delta \varphi_{b1}, \delta \varphi_{a2}, \delta \varphi_{b2} \right)  
\right| ,
\label{eq:criticalI}
\end{align}
where one needs to replace all phase differences $\delta \varphi_\alpha$ by their expression, Eqs.~\eqref{eq:deltaphia1}-\eqref{eq:deltaphib2}, prior to evaluating the maximum.


\begin{figure}[tb]
\centering
\includegraphics[scale=0.72]{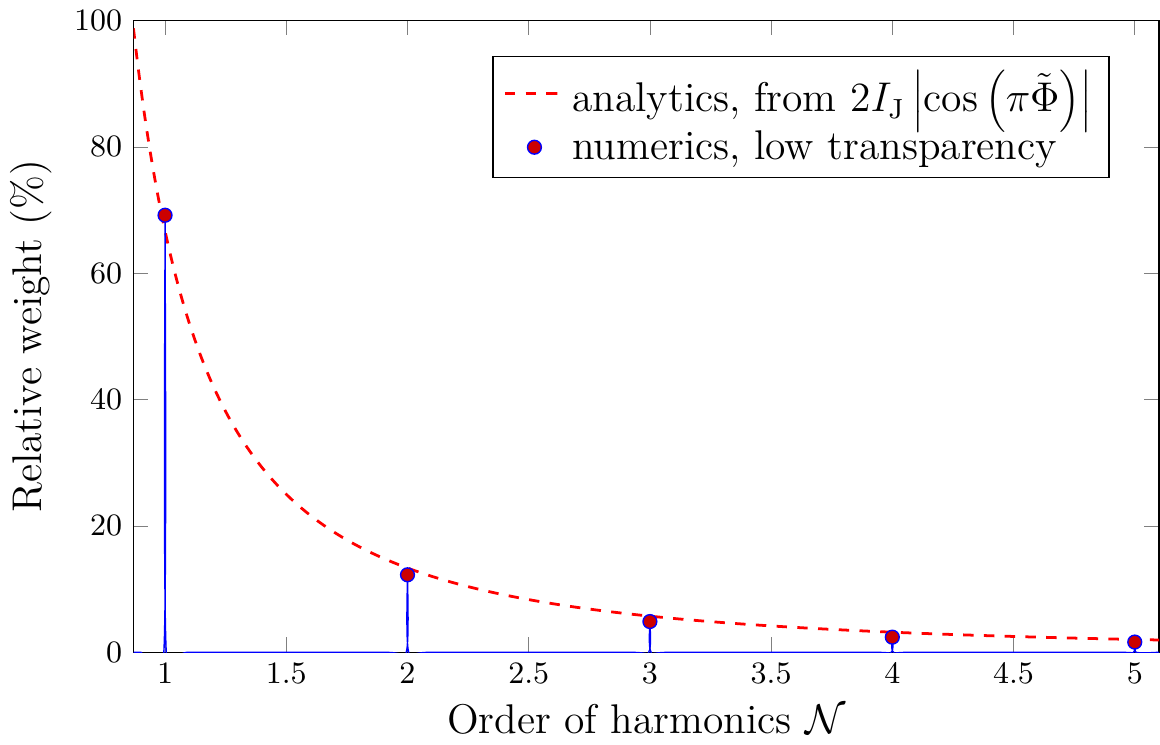}
\caption{The relative weight of the numerically obtained Fourier components (in $\mathcal{N}$ space) of the critical current in the low transparency regime ($\Gamma = 0.01 \Delta$), for identical loops ($\eta=0$) at low temperature, $\beta = 100/\Delta$. This is compared to the analytic solution corresponding to the usual harmonic Josephson current (dotted line). 
}
\label{fig:fourierjosephson}
\end{figure}


\subsection{Low transparency expansion} \label{sec:lowgammaexp}

It is possible at this stage to provide a firmer basis for our earlier phenomenological treatment.
Indeed, starting from Eq.~\eqref{eq:currentvsG}, one can perform a perturbative expansion of the total current $I$ coming out of $S_0$ in powers of the tunneling amplitude.

Focusing on the simpler case of equivalent junctions, i.e. assuming constant tunneling amplitude $t_0$, 
and considering all dot energies to be the same (up to a possible sign, $\epsilon_\alpha = \pm \epsilon$), one can show analytically that the total current takes the form
\begin{align}
\bar{I}_\text{Tot} =& I_{\text{J}} \left[ \sin \delta\varphi_{a1} + \sin \delta\varphi_{b1} + \sin \delta\varphi_{a2} + \sin \delta\varphi_{b2} \right] \nonumber \\
&+ I_{\text{JJ}}  \left[ \sin \left(2 \delta\varphi_{a1}\right) + \sin \left( 2 \delta\varphi_{b1} \right) \right. \nonumber \\
 & \left. \quad + \sin \left( 2 \delta\varphi_{a2}\right) + \sin \left( 2 \delta\varphi_{b2} \right) \right] \nonumber \\
&+ I_{\text{Q}}  \sin \left( \delta\varphi_{a1} + \delta\varphi_{b1} \right), 
\end{align}
when expanding up to forth order in the tunneling rate $\Gamma = 2 \pi \nu(0) t_0^2$, which amounts to taking into account processes involving up to two Cooper pairs. The various contributions are expressed in terms of the microscopic parameters introduced in Sec.~\ref{sec:hamilt} as
\begin{align}
I_{\text{J}} &= \frac{e}{\beta \hbar} \sum_n \frac{\Delta^2}{(\Delta^2 + \omega_n^2) (\epsilon^2 + \omega_n^2)} \Gamma^2 \simeq  \frac{e}{\hbar} \frac{\Gamma^2}{2 \epsilon} \label{eq:IJlowgamma}
 \\
I_{\text{JJ}} &= \frac{e}{\beta \hbar} \sum_n \frac{-\Delta^4 \Gamma^4}{4 (\Delta^2 + \omega_n^2)^2 (\epsilon^2 + \omega_n^2)^2}  \simeq - \frac{e}{\hbar} \frac{\Gamma^4}{16 \epsilon^3}  \\
I_{\text{Q}} &= \frac{e}{\beta \hbar} \sum_n \frac{- \Delta^4 \Gamma^4 }{2 (\Delta^2 + \omega_n^2)^2 (\epsilon^2 + \omega_n^2)^2} \overline{\cos^2 (k_F R) e^{-2 R/\xi}} \nonumber \\
 &\simeq - \frac{e}{\hbar} \frac{\Gamma^4}{16 \epsilon^3} e^{-2 R_0/\xi_0}
\label{eq:IQlowgamma}
\end{align}
where we only kept the leading order contribution to $I_J$ (discarding the third and forth order corrections) and provided simplified forms valid in the large-gap, low-temperature limit $\Delta \gg \epsilon \gg \Gamma, \beta^{-1}$. At this level of approximation, the quartet component is of the same order of magnitude as the Josephson second harmonics (up to the nonlocal prefactor in $R_0$) and does not depend on the specific arrangement of the dots energy levels ($I_\text{Q}$ is unchanged whether $\epsilon_{a1} = \epsilon_{b1}$ or $\epsilon_{a1} = - \epsilon_{b1}$). These properties are specific to the present perturbative treatment and generally not expected to remain valid once one includes contributions from all orders in tunneling. 
The result of Eq.~\eqref{eq:IQlowgamma} also reveals that the amplitude of the quartet current is negative. This $\pi-$type behavior has been observed in Ref.~\onlinecite{jonckheere2013} for multipair dc resonances in an out-of-equilibrium voltage-biased bijunction. It can be attributed to the internal structure of a Cooper pair via the antisymmetry of its wavefunction.

Interestingly, there is no trace of the pair cotunneling current in this derivation. This is an artifact of the low-order expansion, combined with the symmetry of the limit considered here. Indeed, a pair cotunneling current of the order of $I_\text{Q} \times \left( \frac{\epsilon}{\Delta} \right)^2$ does appear, arising from both $S_a$ and $S_b$, but these two terms end up contributing to the total current with an opposite sign.

This calculation not only validates the phenomenological form proposed in Sec.~\ref{sec:setup}, but also justifies that $I_\text{J} \gg |I_\text{JJ}|, |I_\text{Q}|, |I_\text{PC}|$, as assumed in our discussion of the critical current.


\section{Critical current} \label{sec:critical}

We now make use of the general expression \eqref{eq:criticalI} of the critical current of the device in order to explore its evolution for a broad range of parameters, going beyond the low transparency regime.

As argued from our simple phenomenological treatment presented in Sec.~\ref{sec:setup}, we expect quartets (and more generally all multipair processes) to lead to specific signatures in the critical current of the biSQUID, identifiable through their periodicity in the magnetic flux. In what follows, we thus focus on the Fourier spectrum of the critical current.

\subsection{Signatures in Fourier space} \label{sec:fourier}

The critical current can be written in a very general way in terms of its harmonics in $\Phi_\mathcal{A}$ and $\Phi_\mathcal{B}$, namely 
\begin{align}
I_c \left( \Phi_{\mathcal{A}} , \Phi_{\mathcal{B}} \right) = \sum_{n=-\infty}^{+\infty}  \sum_{m=-\infty}^{+\infty} I_{n,m} e^{i 2 n \pi \frac{\Phi_\mathcal{A}}{\Phi_0}} e^{i 2 m \pi \frac{\Phi_\mathcal{B}}{\Phi_0}} .
\end{align}
While we could study independently the variations of the critical current with respect to $\Phi_\mathcal{A}$ and $\Phi_\mathcal{B}$, it makes more sense at this stage to introduce a new set of variables. Indeed, in practice, the magnetic flux through the device is provided by an external homogeneous magnetic field, which affects both $\Phi_\mathcal{A}$ and $\Phi_\mathcal{B}$ in a correlated way. The magnetic fluxes  through loops $\mathcal{A}$ and $\mathcal{B}$ are therefore proportional to one another and only differ as a result of the different surface area enclosed by each loop, so that one can write 
\begin{align}
\Phi_{\mathcal{A},\mathcal{B}} &=  \Phi \left( 1 \mp \eta \right) ,
\end{align}
where we introduced the average magnetic flux $\Phi$ and an asymmetry parameter $\eta$ defined as
\begin{align}
\Phi &= \frac{\Phi_\mathcal{A} + \Phi_\mathcal{B}}{2} \\
\eta&= \frac{\mathcal{S}_\mathcal{B} - \mathcal{S}_\mathcal{A}}{\mathcal{S}_\mathcal{B} + \mathcal{S}_\mathcal{A}} .
\end{align}

In terms of these new variables, the expansion of the critical current $I_c$ takes the form
\begin{align}
I_c \left( \Phi \right) &= \sum_{n=-\infty}^{+\infty}  \sum_{m=-\infty}^{+\infty} I_{n,m} e^{2 i \pi \frac{\Phi}{\Phi_0} \left[ n(1-\eta) + m (1+\eta) \right]} .
\end{align}
The most convenient way to probe for these harmonics is to introduce the Fourier transform with respect to the average phase accumulated around the loops $\varphi = 2 \pi \Phi / \Phi_0$, which is defined as
\begin{align}
\tilde{I}_c \left( \mathcal{N} \right) &= \frac{1}{\Phi_0}  \int d \Phi  ~ I_c \left( \Phi \right)  e^{-2 i \pi \mathcal{N} \frac{\Phi}{\Phi_0}} \nonumber \\
&=   \sum_{n=-\infty}^{+\infty}  \sum_{m=-\infty}^{+\infty} I_{n,m} \delta \left( n(1-\eta) + m (1+\eta) - \mathcal{N} \right) .
\end{align}
It follows that the Fourier-transformed critical current $\tilde{I}_c (\mathcal{N})$ is given by a set of peaks, located at $\mathcal{N} = m + n + (m - n) \eta$, whose height is related to the amplitude of the $(n,m)$ harmonics. Identifying which of these features are present in the spectrum provides a direct and clear indication of what processes are at play in the biSQUID setup. Since the critical current is real and even in $\Phi$, so is its Fourier transform. The Fourier spectrum is thus even in $\mathcal{N}$, and we focus only on positive $\mathcal{N}$.

As a first step, we consider the simplifying approximation of identical junctions. This gives us an opportunity to discuss the typical signatures observed and dress up a physical picture, without obscuring it by dealing with too many parameters.


\begin{figure*}[tb]
\centering
\includegraphics[scale=0.94]{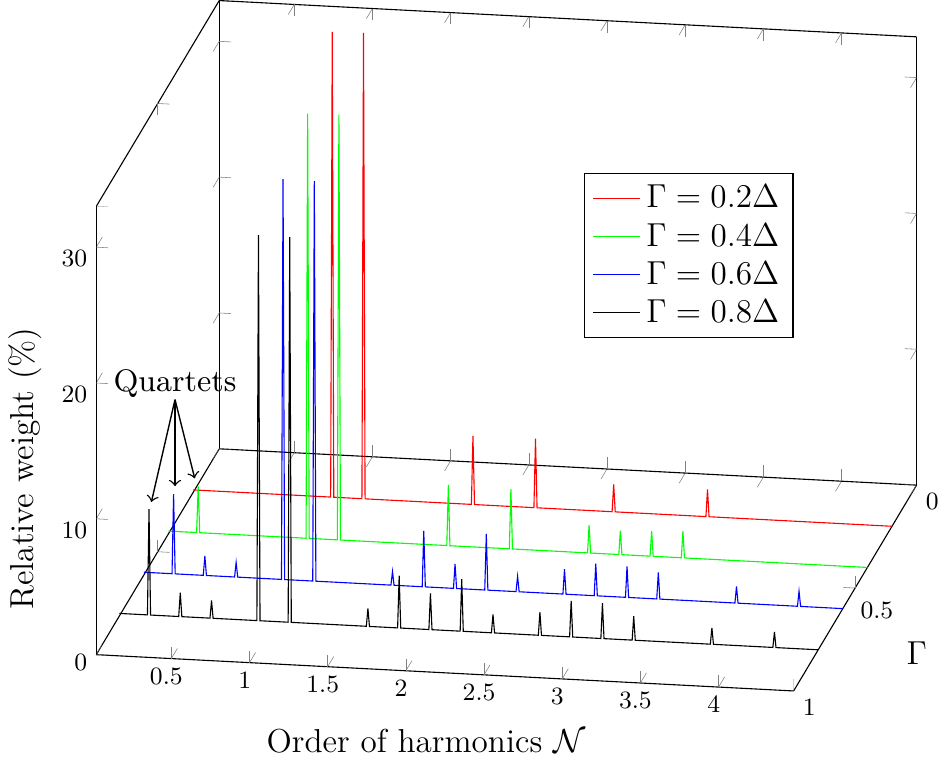}
\includegraphics[scale=0.64]{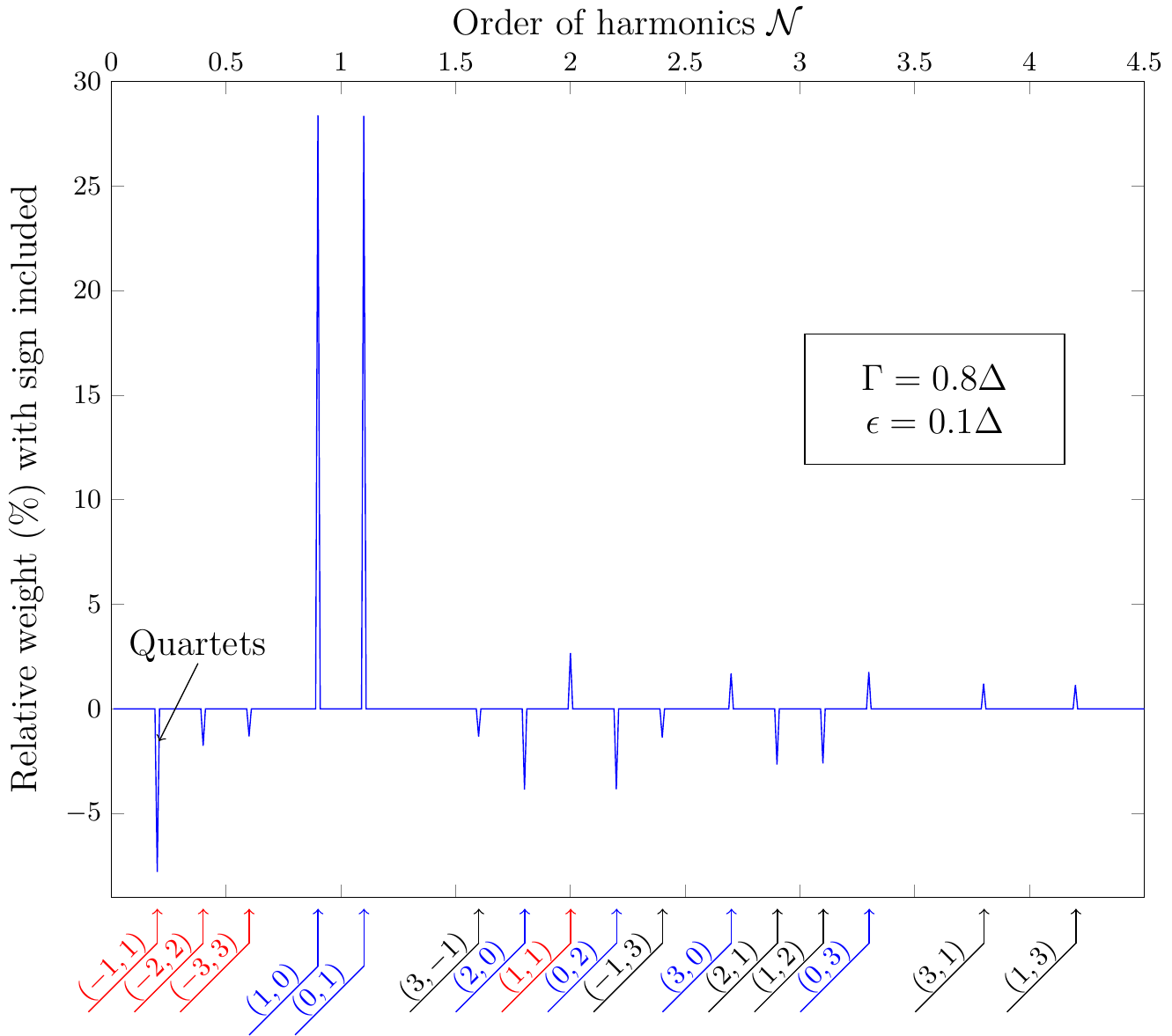}
\caption{(Left) Relative weight of the absolute-valued Fourier components (in $\mathcal{N}$ space) of the critical current as a function of $\Gamma$, for a symmetric arrangement of the energy levels $\epsilon_\alpha = \epsilon = 0.1 \Delta$ (and $\eta=0.1, \beta=100/\Delta, e^{-R_0/\xi_0}=0.9$). (Right) Same relative weight in the specific case of high transparency, $\Gamma = 0.8 \Delta$. There we also restored the sign of the various Fourier amplitudes and identified those in terms of a pair of integers $(n,m)$ corresponding to the periodicity $n \Phi_\mathcal{A}+m\Phi_\mathcal{B}$ of the critical current. 
}
\label{fig:all-symmetric}
\end{figure*}


\subsection{Identical junctions}  \label{sec:vsPhi}

We treat here the case of identical junctions, characterized by the same dot-lead tunneling rate $\Gamma$. For simplicity, we assume that the energy levels $\epsilon_\alpha$ of the dots can only take two values, either $+\epsilon$ or $-\epsilon$, and distinguish various scenarios, based on the possible arrangements of these levels.

We focus on the low-temperature regime ($\beta = 100/\Delta$) and set the attenuation factor $e^{-R_0/\xi_0} = 0.9$, compatible with current experimental realizations.

\subsubsection{Non-resonant case}

We consider first the non-resonant case, for which the energy levels of the dots are non-zero, i.e. detuned with respect to the chemical potential in the superconducting leads. The results are summarized in Fig.~\ref{fig:all-symmetric}.

In particular, we focused on two situations: (a) a symmetric arrangement of the energy levels of the dots involved in the two loops (all chosen equal, $\epsilon_\alpha = \epsilon$) and (b) an antisymmetric arrangement of the energy levels between the two loops (where $\epsilon_{a1}=\epsilon_{a2} = \epsilon$ while $\epsilon_{b1}=\epsilon_{b2} = - \epsilon$). Surprisingly, the obtained Fourier profiles are strictly identical for the symmetric and antisymmetric cases, signaling that the $r-$averaged critical current is insensitive to the arrangement of the dots energy levels, even beyond the low-transparency regime for which we could establish this property semi-analytically. The energy level symmetry has no influence on the Fourier spectrum, which one might interpret as a consequence of some kind of induced particle-hole symmetry, due to the proximity effect from the neighboring superconducting electrodes the dots are in contact with. In particular this means that the quantum dots no longer play the role of an energy filter, as they typically do in hybrid Normal metal - Superconductor - Normal metal structures, such as the Cooper-pair beam splitter,\cite{hofstetter2009} where it allows to favor crossed Andreev reflection over pair cotunneling processes.

In the low transparency regime, the nonlocal effects are not strong enough to be readily identified and one expects the critical current to be insensitive to the arrangement of the energy levels. Indeed, in both cases, the Fourier profile is largely dominated by the usual Josephson current associated to two independent loops. The Fourier spectrum is composed of two sets of peaks, located at multiples of $1-\eta$ and $1+\eta$ respectively. These correspond to harmonics in $\Phi_\mathcal{A}$ and $\Phi_\mathcal{B}$ (of order $(n,0)$ and $(0,m)$ respectively) and can be readily associated with the usual Josephson current flowing through loops $\mathcal{A}$ and $\mathcal{B}$ in absence of any nonlocal coupling between the two. Indeed, in the case of independent loops, one expects  the critical current to be of the form $2 I_\text{J} \sum_j \left| \cos \left( \pi \tilde{\Phi}_j \right) \right|$. This, in turn, leads to a characteristic evolution of the Fourier profile, which we could compare to the one obtained in the present low-transparency  regime. In Fig.~\ref{fig:fourierjosephson}, we show that the weights associated with the different harmonics follow the predicted behavior, confirming that the critical current is dominated by the usual local harmonic Josephson component.


\begin{figure*}[tb]
\centering
\includegraphics[scale=0.94]{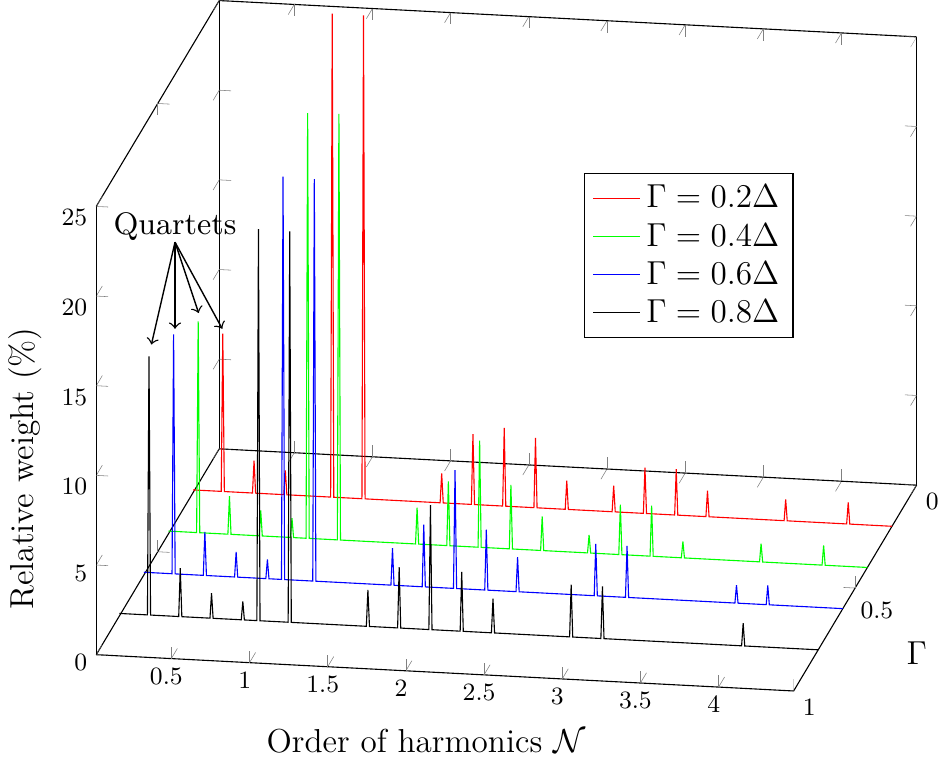}
\includegraphics[scale=0.64]{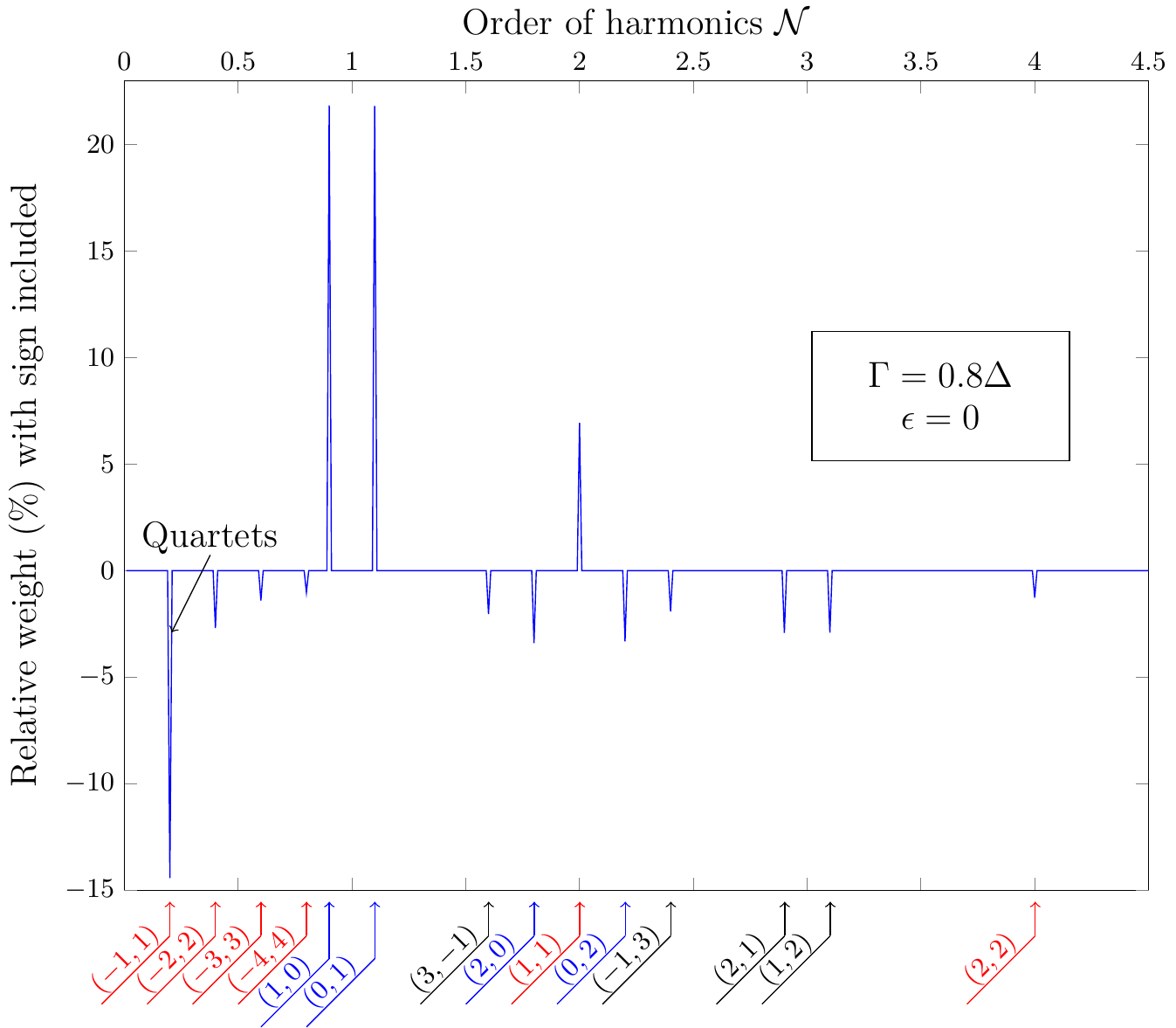}
\caption{(Left) Relative weight of the absolute-valued Fourier components (in $\mathcal{N}$ space) of the critical current as a function of $\Gamma$, for a resonant arrangement of the energy levels $\epsilon_\alpha = \epsilon = 0$ (and $\eta=0.1, \beta=100/\Delta, e^{-R_0/\xi_0}=0.9$). (Right) Same relative weight in the specific case of high transparency, $\Gamma = 0.8 \Delta$. There we also restored the sign of the various Fourier amplitudes and identified those in terms of a pair of integers $(n,m)$ corresponding to the periodicity $n \Phi_\mathcal{A}+m\Phi_\mathcal{B}$ of the critical current. 
}
\label{fig:all-resonant}
\end{figure*}


As the transparency is increased, new structures appear in addition to the ones already observed in multiples of $1 \pm \eta$ (see Fig.~\ref{fig:all-symmetric}). Observing components which involve the fluxes associated with each loop (i.e. with both $n$ and $m$ nonzero) signals the presence of nonlocal processes contributing to the critical current. They are relatively faint for low values of $\Gamma$, but increase as one goes into the high transparency regime. For the highest value of the tunneling amplitude considered here ($\Gamma = 0.8 \Delta$) these nonlocal contributions represent an appreciable total weight of about $30\%$ of the signal.

Out of these new features, the most pronounced one -- a peak in $\mathcal{N}=2 \eta$ -- corresponds to a component of the $r-$averaged critical current with a periodicity $(n=-1,m=1)$ in the fluxes, i.e. periodic in $\Phi_\mathcal{A} - \Phi_\mathcal{B}$. Since it depends on both fluxes, this contribution can only arise from a process involving both the exchange of pairs between $S_0$ and $S_a$ as well as between $S_0$ and $S_b$. Moreover, the periodicity in $\Phi_\mathcal{A} - \Phi_\mathcal{B}$ requires such an exchange to be correlated, and to occur in the same direction, that is from $S_0$ to $S_{a,b}$ or the other way around. This is precisely the microscopic definition of the so-called quartet process, an exchange of two correlated pairs from the central superconducting electrode $S_0$ to $S_a$ and $S_b$. The quartet contribution increases rapidly with $\Gamma$, and becomes substantial in the limit $\Gamma \gg \epsilon$, representing close to $10 \%$ of relative weight, much more than any other components (apart from the leading harmonics in $\Phi_\mathcal{A}$ and $\Phi_\mathcal{B}$). The physical nature of this structure is further confirmed by analyzing the sign of its amplitude (Fig.~\ref{fig:all-symmetric}, right), which turns out to be negative, as expected for the quartet contribution.\footnote{This negative sign is however unrelated to the expected $\pi-$type behavior of the quartet current, as the sign of the Fourier amplitudes of the critical current is not directly related to the sign of the different contributions to the total current.} In the non-resonant case, the biSQUID thus seems a promising candidate to observe signatures of the quartet current, by monitoring the periodicity of the critical current in the total flux, i.e. as a function of the external magnetic field. 

Note that other, much weaker signatures are also present with a periodicity in both fluxes. These could be attributed to either anharmonicities of the quartet current (which would be expected for such high transparency) or to higher order multipair contributions.

It seems that the nonlocal effects are more pronounced when $\Gamma > \epsilon$, which naturally takes us to the regime where this condition is most strongly fulfilled: the resonant case.

\subsubsection{Resonant case}

We now turn to the resonant case, where all energy levels are chosen equal to the chemical potential in the superconducting leads (set to 0 by convention). The results are summarized in Fig.~\ref{fig:all-resonant}.

The low transparency regime is again largely dominated by the local harmonic Josephson current , and absent of any nonlocal signatures. As a result, while the global prefactor $I_\text{J}$ might differ, the Fourier spectrum of the critical current is identical to the one obtained in the non-resonant case.

Contributions associated with nonlocal processes appear rapidly, and already represent about $20\%$ of the signal for tunneling amplitudes as low as $\Gamma = 0.1 \Delta$. As the transparency is further increased, the nonlocal components get stronger, totaling about half of the signal once one reaches $\Gamma \sim \Delta$. 

As in the non-resonant case, there is a proliferation of signatures corresponding to various periodicities in the fluxes $\Phi_\mathcal{A}$ and $\Phi_\mathcal{B}$ (identified through the pair of integers $(n,m)$). The strongest of these nonlocal components is again observed at $\mathcal{N} = 2 \eta$ and corresponds to the quartet current. This contribution grows much more rapidly as a function of $\Gamma$, compared to the non-resonant case, and ends up being twice as large for a comparable value of the tunneling rate. This is consistent with the results obtained in the previous section, as the regime $\Gamma > \epsilon$ proved to be the most suitable to observe signatures of multipair processes, it was to be expected that the resonant case would strengthen this trend further.

Along with the main harmonics $(1,0)$ and $(0,1)$, and the quartet component $(-1,1)$, the Fourier-transformed critical current also shows a substantial contribution for $\mathcal{N} = 2$, corresponding to the periodicity $\Phi_\mathcal{A} + \Phi_\mathcal{B}$ ($n=1, m=1$). Like the quartet component, this contribution also increases with the transparency, and while it was only marginally present in the non-resonant case, it becomes significant at resonance. This feature can be attributed to the pair cotunneling process, which amounts to sending a Cooper pair from $S_a$ to $S_b$ through the central superconducting electrode $S_0$. As argued in Ref.~\onlinecite{freyn2011}, the pair cotunneling process goes hand in hand with the quartet process, and it is not surprising that they appear side by side and behave in a somewhat similar way. As opposed to the low transparency expansion presented in Sec.~\ref{sec:lowgammaexp} where it vanished for symmetry reasons, the pair cotunneling contribution is not negligible here, despite the apparent symmetry of the setup. This reveals the importance of high-order tunneling processes, not accounted for in the  low-$\Gamma$ expansion, which involve the fluxes enclosed in the two loops, thus breaking the symmetry between the two branches of the setup.

The presence of particularly strong quartet signatures makes this resonant case the best candidate to observe such nonlocal contributions to the critical current. We thus continue our study of the resonant case but move on to a more realistic scenario of different junctions in order to determine to what extent our results  are robust.


\begin{figure}[tb]
\centering
\includegraphics[scale=0.6]{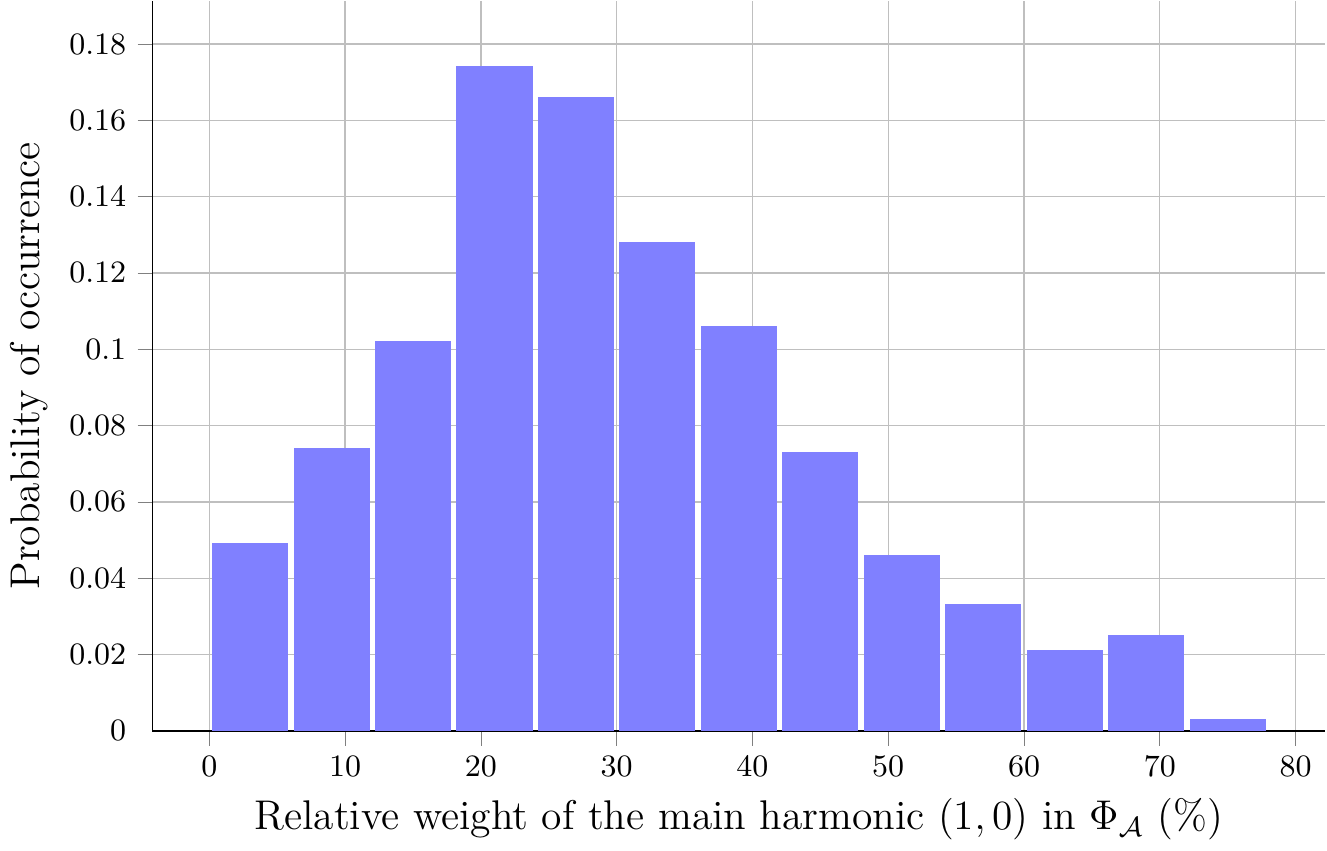}

\includegraphics[scale=0.6]{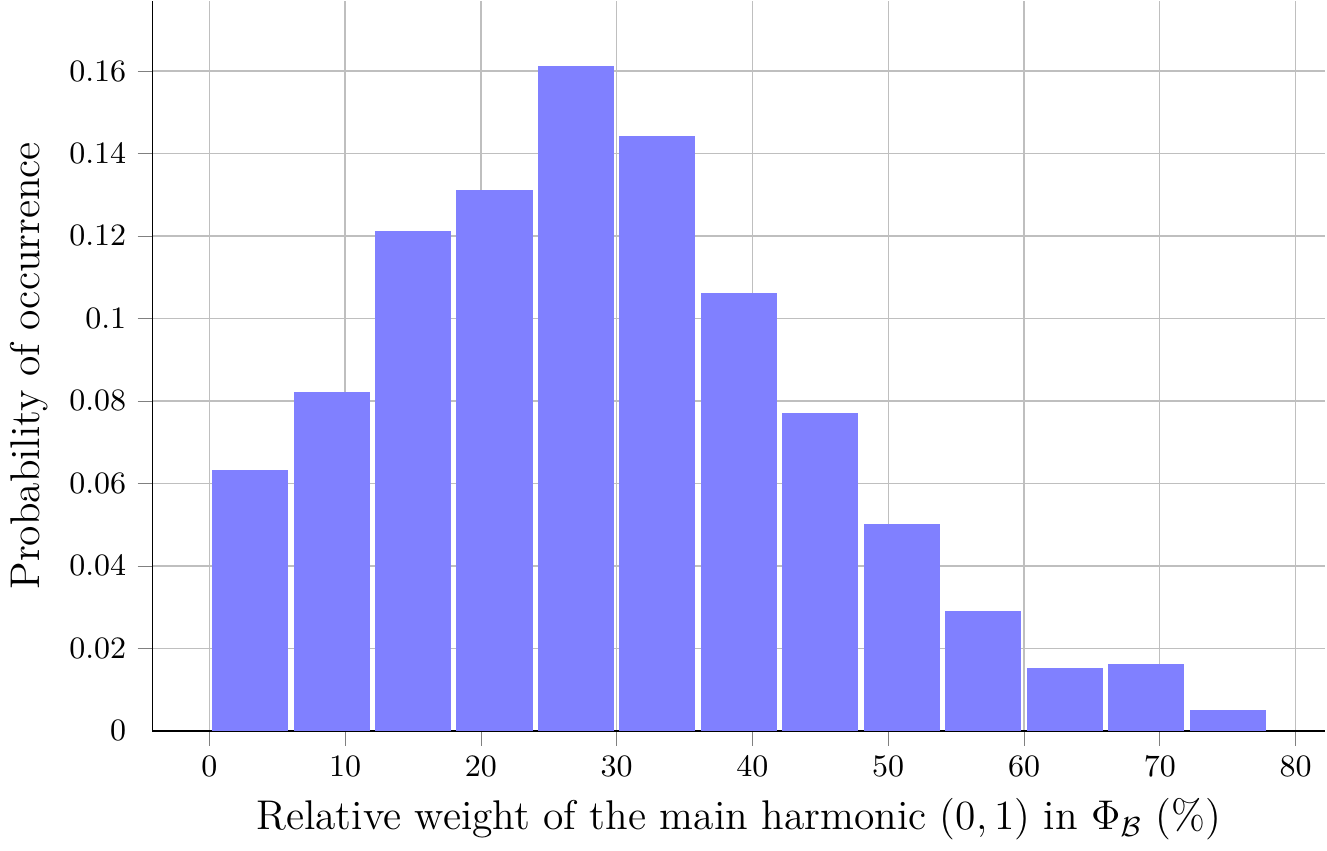}
\caption{Probability of occurrence of the main harmonics and in $\Phi_\mathcal{A}$ (Top) and $\Phi_\mathcal{B}$ (Bottom) as a function of the relative weight they represent in the Fourier spectrum. These were obtained by computing the critical current for 1000 different realizations of the setup with resonant dots ($\epsilon_\alpha = 0$) and randomly chosen tunneling amplitudes $t_{j\alpha}$ between 0 and $\Delta$.
}
\label{fig:histo-local}
\end{figure}


\subsection{Robustness of the results}  \label{sec:robust}

The results of the previous section were obtained in the simpler case of identical junctions, and one might thus argue that they require some specific fine tuning in order to be observed. While the presence of external gates enables a rather precise control of the energy levels of the four quantum dots, the tunneling rates of the junctions depend on the sample design and cannot be manipulated. In particular, making four identical highly transparent Josephson junctions constitutes an experimental challenge. 

\begin{figure}[tb]
\centering
\includegraphics[scale=0.6]{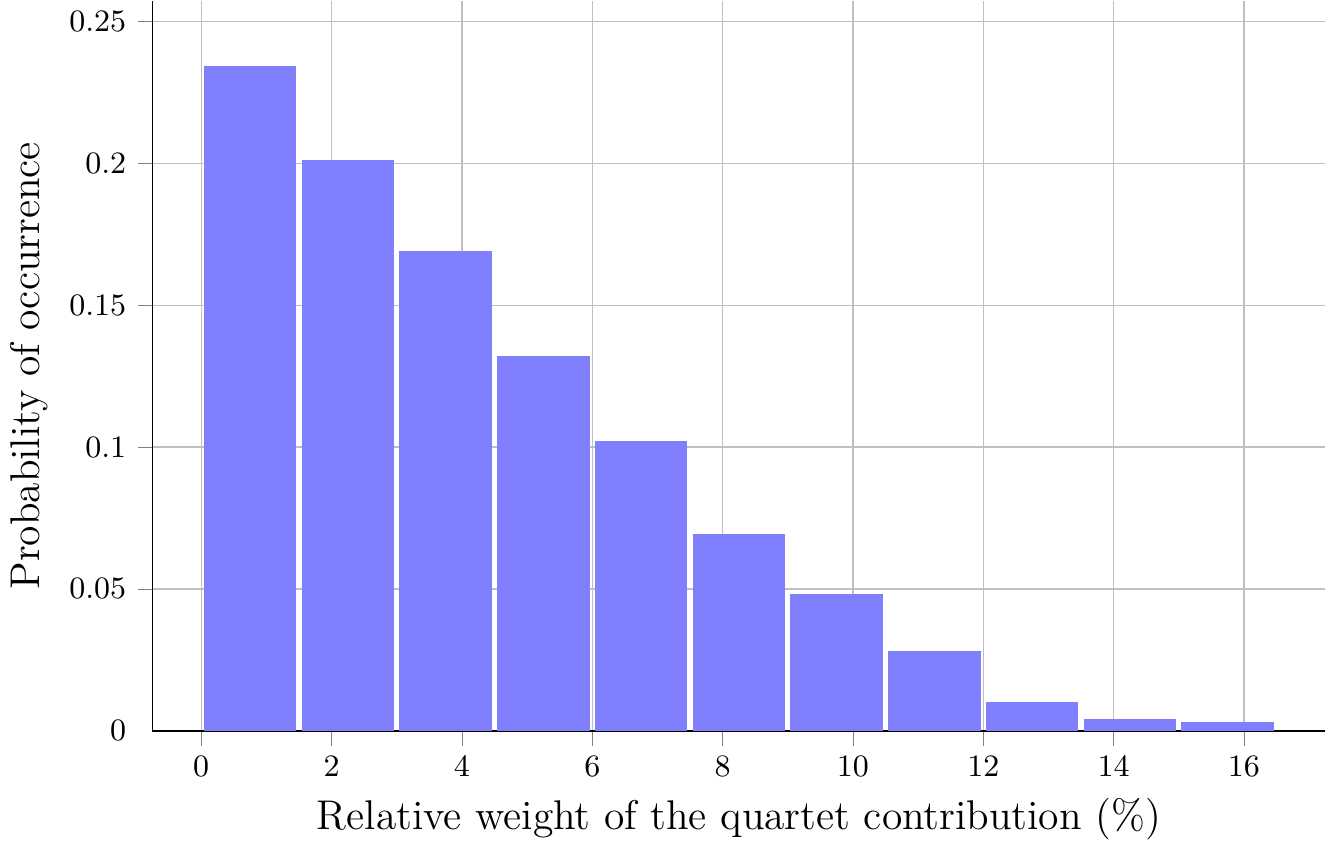}

\includegraphics[scale=0.6]{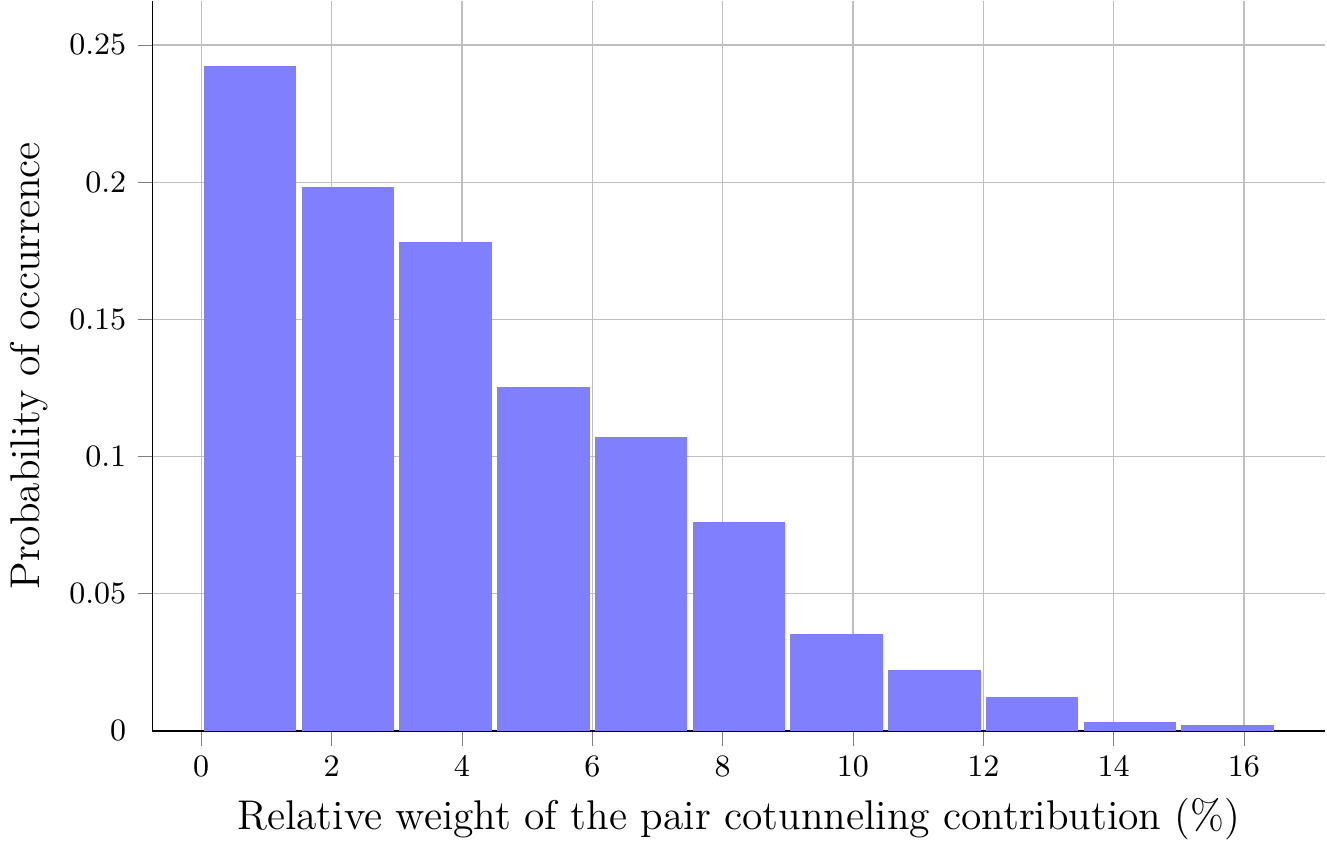}
\caption{Probability of occurrence of the quartet (Top) and pair cotunneling contributions (Bottom) as a function of the relative weight they represent in the Fourier spectrum. These were obtained by computing the critical current for 1000 different realizations of the setup with resonant dots and randomly chosen tunneling amplitudes $t_{j\alpha}$.
}
\label{fig:histo-nonlocal}
\end{figure}

We therefore focus now on a more realistic situation, where we set the dots to be resonant with the superconducting electrodes, but do not make any assumptions concerning the tunneling parameters. To take this even further, we contemplate the possibility of having different tunneling amplitudes $t_{j \alpha}$ for each dot-lead junction. We compute the critical current as a function of the average flux for 1000 different realizations of the setup, each corresponding to a random pick of the eight tunneling amplitudes, chosen following a uniform distribution over the range $ [0 ; \Delta]$. From the resulting Fourier spectra, we monitor the fate of the four most prominent features identified earlier, namely the main harmonics in $\Phi_\mathcal{A}$ and $\Phi_\mathcal{B}$ (corresponding to the features in $(1,0)$ and $(0,1)$), as well as the quartet and pair cotunneling contributions (corresponding to $(-1,1)$ and $(1,1)$ respectively). In Figs.~\ref{fig:histo-local} and \ref{fig:histo-nonlocal}, we show histograms representing the probability of occurrence of each of these contributions as a function of the relative weight they represent in the Fourier spectrum.

From Fig.~\ref{fig:histo-local}, one sees that the main harmonics in $\Phi_\mathcal{A}$ and $\Phi_\mathcal{B}$ follow a very similar distribution. These are clearly the leading contributions to the critical current averaging a relative weight of about $30\%$ each, while together they add up to over $50 \%$ of the signal in about half the realizations.


The results compiled in Fig.~\ref{fig:histo-nonlocal} suggest that the quartet and the pair cotunneling contributions bear some striking similarities when it comes to their weight distribution, further underlining the deep connection between the two processes. Focusing on the quartet component, one readily sees that for a completely random selection of the tunneling amplitudes, the probability of observing a strong quartet signal is rather weak, as less than 60 out of the 1000 realizations show a relative weight beyond $10\%$. However, the quartet signature is more often present than not, its relative weight reaching over $3\%$ for more than half the realizations. On average, the contribution to the critical current associated with quartets culminates a little above $4\%$. To put things into perspective, in the resonant case considered in the previous section, a junction transparency of $\Gamma=0.08 \Delta$ was sufficient to reach the same kind of values, stressing the importance of making the four junctions as identical as possible in practice.

While on average the quartet signature is small, it is still detectable and could be further enhanced by filtering out the main Josephson harmonics in $\Phi_\mathcal{A}$ and $\Phi_\mathcal{B}$ which are easily identifiable. One also has to keep in mind that the situation considered here is among the most unfavorable ones, and while practical realizations should be tailored to promote nonlocal multipair processes, even a random realization still has a solid chance of revealing their specific signatures.


\section{Competing effects} \label{sec:competing}

Various physical phenomena can be responsible for a coupling between degrees of freedom from the two loops, leading to nonlocal effects whose signatures might mask a pure quartet signal. We hereby consider two of the most likely candidates susceptible to appear in our biSQUID device.

\subsection{Mutual inductance}

The setup under consideration is comprised of two loops where current can flow, so that one should take into account the geometrical inductance of such a circuit. While there could be both local and mutual inductances at play in the setup, the latter is the most likely to lead to strong nonlocal signatures as it couples the currents of pairs flowing through junctions $a1$ and $b1$. We could check that the presence of a self inductance only marginally modifies our results and thus decided, for simplicity, to focus on the effect of the mutual inductance alone.

Due to the mutual geometrical inductance $M$ between loops $\mathcal{A}$ and $\mathcal{B}$, the fluxes felt by the electrons circulating in the circuit is no longer set by the external magnetic field alone. Instead, there exists an additional contribution on top of this external flux (hereby labeled $\Phi^{\text{ext}}_{\mathcal{A},\mathcal{B}}$) which depends on the current flowing in the nearby loop. Following the conventions introduced in Fig.~\ref{fig:setup} and Eqs.~\eqref{eq:deltaphia1}-\eqref{eq:deltaphib2} for the currents and fluxes, we write
\begin{align}
\label{eq:newPhiA}
\Phi_\mathcal{A} &= \Phi^{\text{ext}}_\mathcal{A} - M \left( I_{b1} - I_{b2} \right) \\
\Phi_\mathcal{B} &= \Phi^{\text{ext}}_\mathcal{B} - M \left( I_{a2} - I_{a1} \right) ,
\label{eq:newPhiB}
\end{align}
where the mutual inductance $M$ is positive.

\begin{figure}[tb]
\centering
\includegraphics[scale=0.9]{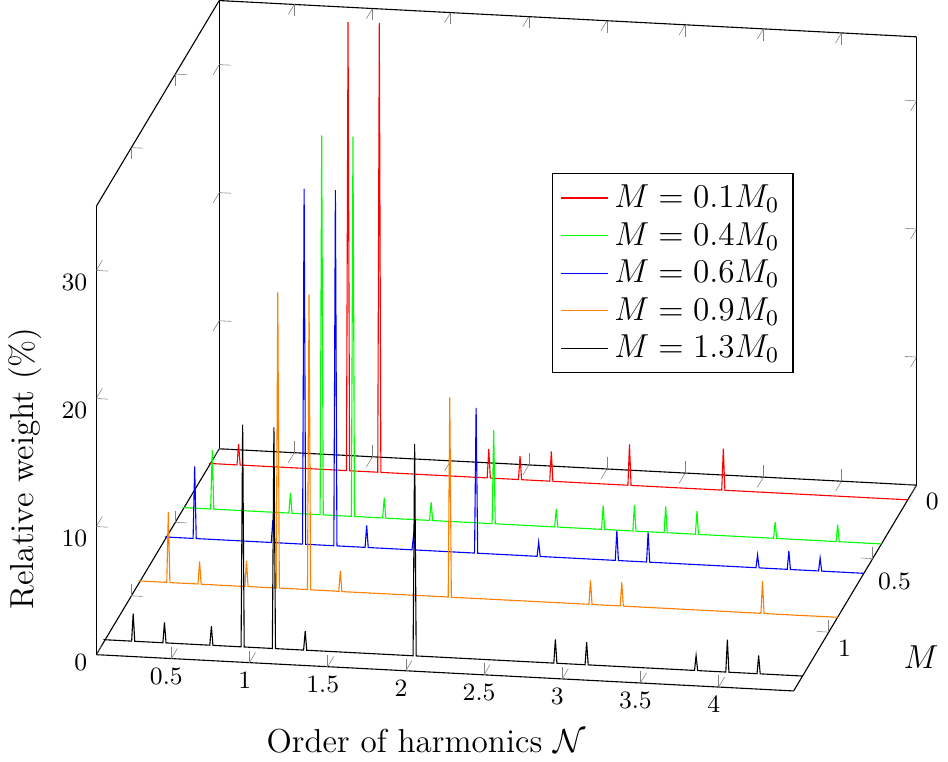}

\caption{Relative weight of the Fourier components of the critical current as a function of the mutual inductance $M$ between loops (expressed in units of $M_0 = \hbar^2/(e^2 \Delta)$), in the resonant ($\epsilon=0$) highly transparent regime ($\Gamma=0.8 \Delta$). All other parameters have been chosen identical to Figs.~\ref{fig:all-symmetric} and \ref{fig:all-resonant}, apart from the attenuation factor, $e^{-R_0 / \xi_0} = 0$. 
}
\label{fig:mutualinductance}
\end{figure}

Since some of the junction currents $I_\alpha$ depend explicitly on the fluxes, it is obvious from Eqs.~\eqref{eq:newPhiA}-\eqref{eq:newPhiB} that the total current through the device now has to be computed  self-consistently. Performing this self-consistent treatment in addition to the $r-$averaging and on top of the maximization (in order to extract the critical current) makes the numerical calculation a lot more demanding. In order to estimate the competing effects associated with the mutual inductance, we thus consider the simpler case of two disconnected loops (i.e. without any microscopic nonlocal coupling) and turn on the mutual inductance progressively, monitoring the Fourier-transformed critical current. The results are provided in Fig.~\ref{fig:mutualinductance}, where for illustrative purposes, we focused on the resonant highly transparent regime.

Whereas there are no noticeable modifications of the Fourier spectrum for weak values of the mutual inductance, new signatures start appearing beyond $0.1 M_0$ (where $M_0 = \hbar^2/(e^2 \Delta)$), a threshold which tends to increase as the tunneling rate $\Gamma$ is reduced (not shown).

These signatures grow rapidly with $M$ and are mainly located at $\mathcal{N} = 2$ and $\mathcal{N} = 2 \eta$ (i.e. the $(1,1)$ and $(-1,1)$ components respectively). Unlike the general situation considered in the previous section, the most pronounced component involving both fluxes corresponds here to $(1,1)$ rather than $(-1,1)$, a characteristic specific to the presence of mutual inductance, which could be used in actual experiments to detect the presence of such effects. Indeed, this can be explained by noticing that the mutual inductance tends to {\it anticorrelate} the currents in branches $a1$ and $b1$ in order to decrease the total current in the central electrode. On the contrary, the quartet mechanism tends to {\it correlate} the currents $I_{a1}$ and $I_{b1}$ reinforcing the current in the central electrode. The antagonistic effects of quartets and mutual inductance allows to discriminate both mechanisms. For the largest value of the mutual inductance considered here, this structure at $\mathcal{N} = 2$ even becomes the leading component of the Fourier signal, overcoming the contributions which depend on only one of the two fluxes. While a peak at $\mathcal{N} = 2 \eta$ is visible (and as such, could be mistaken with signatures from the quartet process), it stays relatively weak ($5\%$ at best) and even decreases back for large values of $M$.

In the end, although mutual inductance leads to structures in the critical current located at the same position in $\mathcal{N}$ space, the behavior of these new features is specific enough to not be confused with quartet and pair cotunneling processes. Still, the mutual inductance needs to be carefully estimated in experimental setups\footnote{For a gap energy $\Delta \sim 100~\mu$eV, the mutual inductance considered here is between 50 and 500 pH, which should be a reasonable order of magnitude of what is realized in actual setups.} and taken into account when attempting to fit the data.


\subsection{Direct tunneling}

Although absent from nanowire-based systems, direct tunneling between quantum dots is known to occur in carbon nanotubes. This generates a new contribution to the bare dot Hamiltonian, of the form
\begin{align}
\mathcal{H}_{D,\text{direct}} = t_d {\bf d}^\dagger_{a1}  \sigma_z  {\bf d}_{b1} + \text{H.c.}.
\end{align}
where $t_d$ is the direct tunneling amplitude between dots $a1$ and $b1$. This term opens a new channel for current to flow between the two loops, competing with the nonlocal exchange of pairs through $S_0$. 

\begin{figure}[tb]
\centering
\includegraphics[scale=0.9]{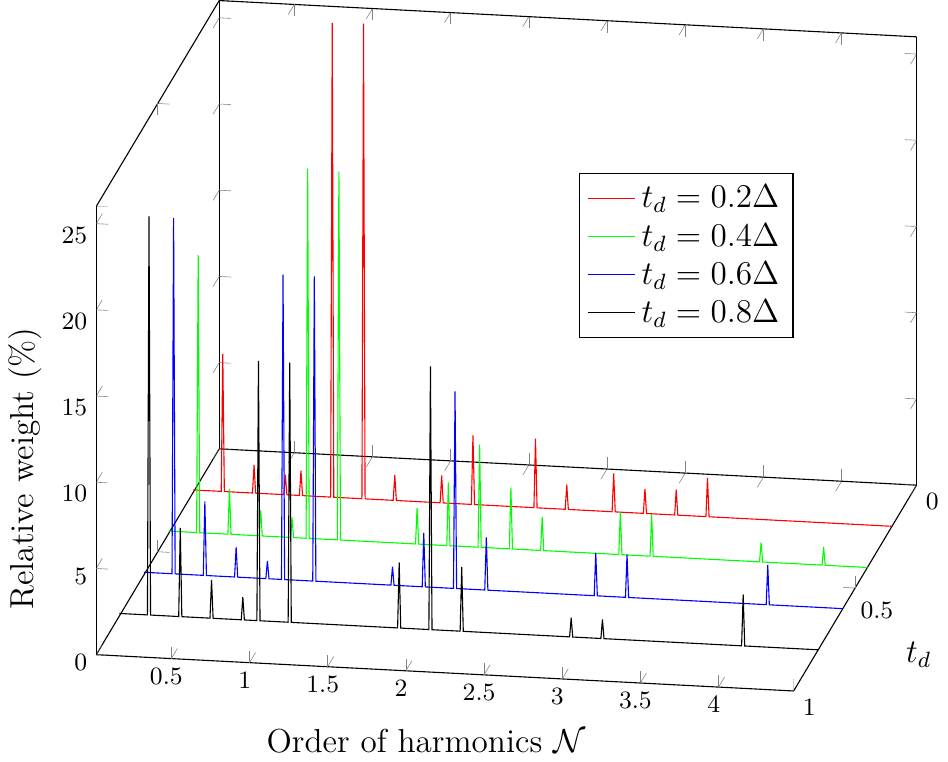}

\caption{Relative weight of the Fourier components of the critical current as a function of the direct tunneling amplitude $t_d$, in the resonant ($\epsilon=0$) highly transparent regime ($\Gamma=0.8 \Delta$). All other parameters have been chosen identical to Figs.~\ref{fig:all-symmetric} and \ref{fig:all-resonant}, apart from the attenuation factor, $e^{-R_0 / \xi_0} = 0$. 
}
\label{fig:directtunneling}
\end{figure}

To get a flavor of the effect of direct tunneling, let us compute the critical current as a function of $t_d$ in the simpler case where the nonlocal coupling through the central superconducting electrode has been turned off. The new term that now appears in the total Hamiltonian changes the form of the Matsubara Green's function for the dot electrons by affecting its anomalous part $f(i\omega_n)$, which becomes
\begin{align}
f (i \omega_n) &= 
\begin{pmatrix}
- t_d & 0  \\
0 & t_d  \\
\end{pmatrix}.
\label{eq:anomalouswithtd}
\end{align}
The critical current is then obtained from Eqs.~\eqref{eq:totalI} and \eqref{eq:criticalI} as before. The Fourier-transformed critical current as a function of the direct tunneling $t_d$ is presented in Fig.~\ref{fig:directtunneling} for the resonant highly transparent case.

As it turns out, direct tunneling can lead to signatures that are very similar to nonlocal multipair processes. In the resonant case considered here, even for rather small values of $t_d$ (say, $0.1 \Delta$) new peaks appear  in the Fourier spectrum corresponding precisely to the contributions of interest, namely $(-1,1)$ and $(1,1)$ (associated earlier to quartet and pair cotunneling processes). Moreover, these features grow rapidly as $t_d$ is increased, and for a typical value of only $t_d = 0.4 \Delta$, the resulting Fourier-transformed critical current looks almost identical to the one obtained in Fig.~\ref{fig:all-resonant} in presence of nonlocal crossed Andreev reflection (same set of harmonics with similar weights, only the ones in the vicinity of $\mathcal{N}=4$ differ). Indeed, from inspecting the anomalous Green's functions in presence of a coupling through $S_0$, Eq.~\eqref{eq:anomalousGreen}, and in presence of direct tunneling, Eq.~\eqref{eq:anomalouswithtd}, one readily sees that $t_d$ roughly assumes the same role as the prefactor $\Gamma e^{-R_0 / \xi_0}/2$ (up to complications related to $r-$averaging). Indeed, direct interdot tunneling offers an alternative channel for the production of correlated pairs, as shown in Refs.~\onlinecite{herrmann2010, burset2011}. Yet, it is interesting to discriminate this mechanism from the quartet one, for instance by synchronous detection while slowly varying one key parameter of the setup.


\section{Summary and conclusion} \label{sec:conclusion}

To summarize, our main results are as follows:

(a) the phenomenological model is sufficient to capture the main ingredients of the physics involved in this system in the low transparency regime;

(b) studying the flux-dependence of the critical current through a full microscopic calculation leads, for identical junctions in the non-resonant regime, to new signatures. These could be attributed to nonlocal multipair processes;

(c) in the case of resonant dots, and for highly transparent junctions, these quartet signatures become particularly strong (of the same order as the Josephson contribution) making this regime the most promising to investigate multipair production;

(d) our results are robust against strong variations of tunneling amplitude throughout the setup. Even in the worst case scenario of 8 very different tunneling parameters, there is a strong probability of observing a quartet signal;

(e) the most likely competing effects either lead to qualitatively different signatures or open a new channel that contributes to the formation of correlated pairs, in addition to the standrad crossed Andreev reflection.

This work thus provides an alternative experimental setup for the observation of nonlocal multipair processes. Unlike previous work on a voltage-biased all-superconducting bijunction\cite{jonckheere2013}, the current study focuses on an equilibrium situation, in the coherent dissipationless regime. Making use of interferometry, this allows for the observation of specific features associated with quartet processes that do not require to study intricate phase and voltage dependences, and hopefully provides an unequivocal experimental signature of these phenomena.

In conclusion, we proposed a setup consisting of four Josephson junctions, defining two twinned loops in a nanotube (or nanowire) based system connected to three superconducting electrodes, which we dubbed a biSQUID. We presented a phenomenological argument to motivate our analysis then derived a microscopic theory allowing a careful description of the setup in the full parameter space. We showed that by measuring the critical current as a function of the average flux through the loops (i.e. the external magnetic field), the device could reveal signatures associated with nonlocal multipair contributions, in particular the so-called quartet process, thus making them experimentally observable in an equilibrium situation. We also suggested that potential experimental realizations pay specific attention to some competing effects which might interfere with these very signatures.

A natural extension of the present work consists in taking into account local Coulomb interaction on the dots. This constitutes a real challenge, which could be tackled through various approximate treatments, such as perturbative diagrammatic resummation or self-consistent mean-field approaches.\cite{rech2012} Interactions are expected to be particularly relevant when dealing with closed dots, with weak tunneling amplitudes to the leads. We showed, however, that the most interesting regime to observe quartet signatures in the biSQUID setup corresponds to the opposite case of large values of the tunneling rate $\Gamma$, for which Coulomb interactions play a more limited role. As long as one stays away from the deep Kondo regime (for which the occupation of the dots is close to 1 at all times), we therefore do not expect any dramatic qualitative change of our results due to Coulomb interactions.

\begin{acknowledgments}
We acknowledge the support of the French National Research Agency, through the project ANR-NanoQuartets (ANR-12-BS1000701). Part of this work has been carried out in the framework of the Labex Archim\`ede ANR-11-LABX-0033 and of the A*MIDEX project ANR-11-IDEX-0001-02 (J.R., T.J. and T.M.). 
The authors acknowledge seminal discussions with T. Kontos concerning the biSQUID scheme.
\end{acknowledgments}

\appendix

\section{Tunneling self-energy} \label{app:sigmaT}

The tunneling self-energy introduced in the text is given, in terms of Matsubara frequency, by
\begin{align}
\Sigma_{\alpha \gamma} (i \omega_n) =& \sum_{j} 
t_{j \gamma} t_{j \alpha}^*  \sigma_z e^{-i \sigma_z \varphi_j /2} \nonumber \\
&\times \left[ \sum_k \left( i\omega_n \mathds{1} - \xi_k \sigma_z - \Delta_j \sigma_x  \right)^{-1} e^{i k (r_{j\alpha}-r_{j\gamma})} \right] \nonumber \\
& \times  \sigma_z e^{i \sigma_z \varphi_j /2} ,
\label{eq:Appsigma}
\end{align}
where we substituted in Eq.~\eqref{eq:tunnelsigma} the expression for the lead electrons Green's function, and the tunneling parameters.

Performing the $k-$integral requires information on the dimensionality of the system. It is believed\cite{burset2011} that the contact between dots and superconducting leads actually occurs within the carbon nanotube, between an electrostatically confined region (the dot) and a proximity-induced superconducting region, which then acts as an effective 1D superconductor. Within this assumption, the integral over momentum reduces to
\begin{align}
\sum_k &\left( i\omega_n \mathds{1} - \xi_k \sigma_z - \Delta_j \sigma_x  \right)^{-1} e^{i k (r_{j\alpha}-r_{j\gamma})}   \nonumber \\
=& \sum_k \frac{-e^{i k (r_{j\alpha}-r_{j\gamma})}}{\omega_n^2+\Delta_j^2+\xi_k^2} \left( i\omega_n \mathds{1} + \xi_k \sigma_z + \Delta_j \sigma_x  \right) \nonumber \\
=& \nu (0) \int d\epsilon \frac{\cos \left( k_F \sqrt{1+\frac{\epsilon}{\mu}} (r_{j\alpha}-r_{j\gamma})\right) }{\epsilon^2 + \omega_n^2 + \Delta_j^2} \nonumber \\
& \qquad \times \left( i\omega_n \mathds{1} + \epsilon \sigma_z + \Delta_j \sigma_x  \right) \nonumber \\
\simeq& \pi \nu (0) e^{-R_{j,\alpha\gamma}/\xi(i\omega_n)} \nonumber \\
& \times \left[  \frac{\cos k_F R_{j,\alpha\gamma}}{\sqrt{\omega_n^2 + \Delta^2}} \left( i\omega_n \mathds{1} + \Delta_j \sigma_x  \right)  - \sin k_F R_{j,\alpha\gamma} \sigma_z   \right] ,
\end{align}
where we introduced $R_{j,\alpha\gamma} = | r_{j\alpha}-r_{j\gamma} |$ as well as $\xi(i\omega_n) = \frac{\xi_0 \Delta}{\sqrt{\Delta^2 + \omega_n^2}}$. Here $\nu(\epsilon)$ is the density of states of the superconducting region, which we assumed to be constant close to the Fermi level.

Substituting this result back into Eq.~\eqref{eq:Appsigma}, one has, for the tunneling self-energy in Matsubara frequency space
\begin{align}
\Sigma_{\alpha \gamma} (i \omega_n) =& \pi \nu (0) \sum_{j} 
t_{j \gamma} t_{j \alpha}^*   e^{-R_{j,\alpha\gamma}/\xi(i\omega_n)} \nonumber \\
& \times \left[  \frac{\cos k_F R_{j,\alpha\gamma}}{\sqrt{\omega_n^2 + \Delta^2}} \left( i\omega_n \mathds{1} - \Delta_j e^{- i \sigma_z \varphi_j} \sigma_x  \right) \right. \nonumber \\ 
& \qquad - \sin k_F R_{j,\alpha\gamma} \sigma_z   \Bigg] .
\end{align}
In practice, the only nonlocal terms allowed by the tunneling amplitudes involve coupling between dots $a1$ and $b1$ through lead $S_0$. Therefore, we only need to introduce one length scale $R_{j,\alpha\gamma}$ corresponding to $R_{0;a1,b1} = R_{0;b1,a1} =R$.

\end{document}